# BCIJelly: An integrated ecosystem for brain–computer interface research

Liyuan Han[1,7,#] , Xinrui Yang[1,#], Tianyu Zheng[1,6,#], Qizhi Yang[3,4], Yitao Qin[1,6], Liang Chen[3,4], Qinglai Wei[3,4], Binjie Hong[1], Xinhe Zhang[1], Rui Xiong[1], Yong Gu [1,5], Mu-ming Poo[1], Bo Xu[3,4], Chengyu Li[2,*], Tielin Zhang[1,4,5,*]

[1] Center for Excellence in Brain Science and Intelligence Technology, Institute of Neuroscience, Chinese Academy of Sciences, Shanghai 200031, China.

[2] Lingang Laboratory, Shanghai 200031, China.

[3] Institute of Automation, Chinese Academy of Sciences, Beijing 100190, China.

[4] School of Artificial Intelligence, University of Chinese Academy of Sciences, Beijing 100049, China.

[5] State Key Laboratory of Brain Cognition and Brain-inspired Intelligence Technology.

[6] School of Advanced Interdisciplinary Sciences, University of ChineseAcademy of Sciences

* Corresponding authors: zhangtielin@ion.ac.cn and tonylicy@lglab.ac.cn

[#] These authors contributed equally to this work

## Abstract

Brain-computer interface (BCI) research relies on multistage computational pipelines, yet progress remains constrained by fragmented data formats, heterogeneous decoder implementations and hardware-specific deployment toolchains, and researchers lack an integrated workflow. Here, we fill this gap with BCIJelly, a unified computational ecosystem that integrates 18 curated BCI datasets, 15 benchmark decoders and an algorithmic library of 80 reusable modules, an automated architecture search (AAS) procedure, and hardware-aware deployment through the *toChip* pipeline within a single Python framework. AAS constructs task-specific decoders without manual architecture design. It is further extended into a closed-loop mode guided by a large language model (LLM), which uses task specifications, module descriptions and search history to support multitask and cross-species decoding. The *toChip* pipeline compiles trained decoders for execution on neuromorphic chips, enabling energy-efficient deployment for BCI systems. An accompanying visualization software provides a graphical interface to the full workflow, making BCIJelly accessible without programming. We validate BCIJelly across five BCI paradigms (motor, visual, speech, emotion and auditory) with recordings from humans, macaques and mice, and single-task, multitask and cross-species decoding settings. BCIJelly establishes a unified and extensible infrastructure that bridges decoder development and hardware-aware deployment for BCI research.

## Introduction

A central goal of brain-computer interface (BCI) research is to decode behavioral intent from neural activity and translate it into actionable control signals, with applications ranging from motor restoration to communication. Recent advances in high-density neural recording and deep learning have substantially expanded what can be decoded from neural populations, with state-of-the-art decoders now spanning motor, visual[1], speech[2-4], emotion and auditory[5] paradigms across multiple species. However, BCI research increasingly suffers from a different kind of bottleneck. The full workflow, from raw neural recordings through decoder development to deployment on specialized hardware, is fragmented across incompatible tools, each tailored to a specific dataset, paradigm or hardware target. As a result, researchers spend substantial effort writing custom infrastructure at each stage rather than focusing on the scientific or engineering questions of interest.

Existing computational tools for BCI research have typically focused on individual stages of this workflow. The established libraries provide standardized data handling and storage (Neo[6] and Neurodata Without Borders[7,8], NWB). Specific decoding algorithms have been developed as standalone implementations, ranging from linear (MLP) and recurrent (LSTM) models to recent attention-based (NDT[9,10] and POYO[11]) and latent variable architectures (LFADS[12], Cycle-GAN[13] and NoMAD[14]). Standardized benchmarks have emerged to compare decoder performance across curated datasets (FALCON[15], Neural Latents Benchmark[16], NLB and OmniEEG-Bench[17]). While each of these efforts addresses an important component of the workflow, combining them into a working BCI pipeline still requires substantial integration effort, and there is no unified framework that combines together preprocessing, decoder evaluation and development, and hardware deployment within a single ecosystem.

More recent libraries have begun to consolidate parts of this workflow. NeuroAI[18],

a recently released Python suite, unifies data processing (NeuralSet), dataset curation (NeuralFetch), model training (NeuralTrain) and benchmarking (NeuralBench) across diverse neural modalities including fMRI, MEG, EEG and spikes, oriented toward linking neural recordings to large pretrained models for Neuro-AI research rather than the decoding and hardware deployment workflow specific to BCI research. TorchBrain[11,19] provides PyTorch-compatible building blocks such as datasets, samplers and modules for designing neural decoders, with reference implementations of recent architectures including POYO[11], but without offering a standardized preprocessing or end-to-end decoding pipeline. While these libraries have substantially advanced deep learning for neural data, three challenges remain particularly central to BCI research. First, no single-decoder architecture is dominant across BCI paradigms or datasets, with the best-performing model often differing from one dataset to the next; thus, researchers must test and tune multiple architectures for each new task. Second, scaling decoder development across tasks would benefit from search procedures that can reuse prior search history, diagnose failed architectures and incorporate newly designed modules, rather than repeatedly exploring a fixed model space from scratch. Third, practical BCI use depends on low-power operation, which motivates deploying trained decoders on efficient edge or neuromorphic hardware[20-24] rather than relying exclusively on general-purpose GPUs. Crucially, neither the NeuroAI nor the TorchBrain addresses any of these needs.

Here, we address these challenges with BCIJelly, a unified computational ecosystem for BCI research that integrates data preprocessing, benchmark decoders, automated architecture search and hardware-aware deployment within a single Python framework. BCIJelly provides 18 curated BCI datasets, and 15 complete decoders representative of the major methodological families used in BCI research, together with a library of 80 reusable modules covering fully connected, convolutional, backbone and attention families. Building on this module library, the AAS procedure automatically samples, assembles and evaluates module combinations to construct task-specific decoders without manual architecture design. We further extend AAS into a closed-loop procedure with an LLM that uses task specifications, module descriptions and search history to plan decoder searches, diagnose failed candidates and design new modules for multitask settings. Trained decoders are then compiled through the *toChip* pipeline for execution on neuromorphic chips, with substantial reductions in power consumption compared with general-purpose GPUs. An accompanying visualization platform provides a graphical interface to the full workflow, enabling exploration of neural recordings and decoding results without programming. Validation across five BCI paradigms, three species and single-task, multitask and cross-species decoding settings demonstrates that BCIJelly can support diverse BCI workflows.

# Results

## Overview of the BCIJelly computational ecosystem

BCIJelly is a unified computational ecosystem that combines data processing, model development, hardware deployment and visualization within a single framework (**Fig. 1a**). The framework follows a modular design: Each stage of the workflow is implemented as an independent component that follows a unified input–output format (i.e., API), so that modules can be combined, swapped or extended on demand. On the data side, the framework standardizes heterogeneous BCI datasets across five paradigms (motor, visual, speech, emotion and auditory), and supports both classification and regression analyses (**Fig. 1a1**).

At its core, BCIJelly provides a modular model library with two levels: a collection of algorithm-level decoders (LSTM, Transformer, LFADS, Cycle-GAN and others) that serve as ready-to-use baselines, and a pool of module-level components (convolutional, residual, attention, full connection and others) that can be reused to compose new architectures, all of which are exposed through a unified API (**Fig. 1a2**). Drawing from the module pool, AAS programmatically samples, assembles and evaluates module combinations to generate entirely new, task-specific decoders, automating much of the architecture design process.

Models trained on GPUs are compiled through a hardware-aware pipeline and deployed onto neuromorphic chips such as TaiBai and Lynxi HE200, with FPGAs serving as presilicon prototypes during chip development (**Fig. 1a3**). Beyond the core pipeline, BCIJelly also includes interactive visualization software that provides a single graphical interface for data import and export, signal visualization, electrode localization and data analysis (**Fig. 1a4**).

A representative workflow built with the *bcijelly* Python library shows how the entire pipeline, from loading a dataset to training a decoder, converting it for neuromorphic deployment and visualizing the signals, can be expressed in just a few lines of code (**Fig. 1b**).

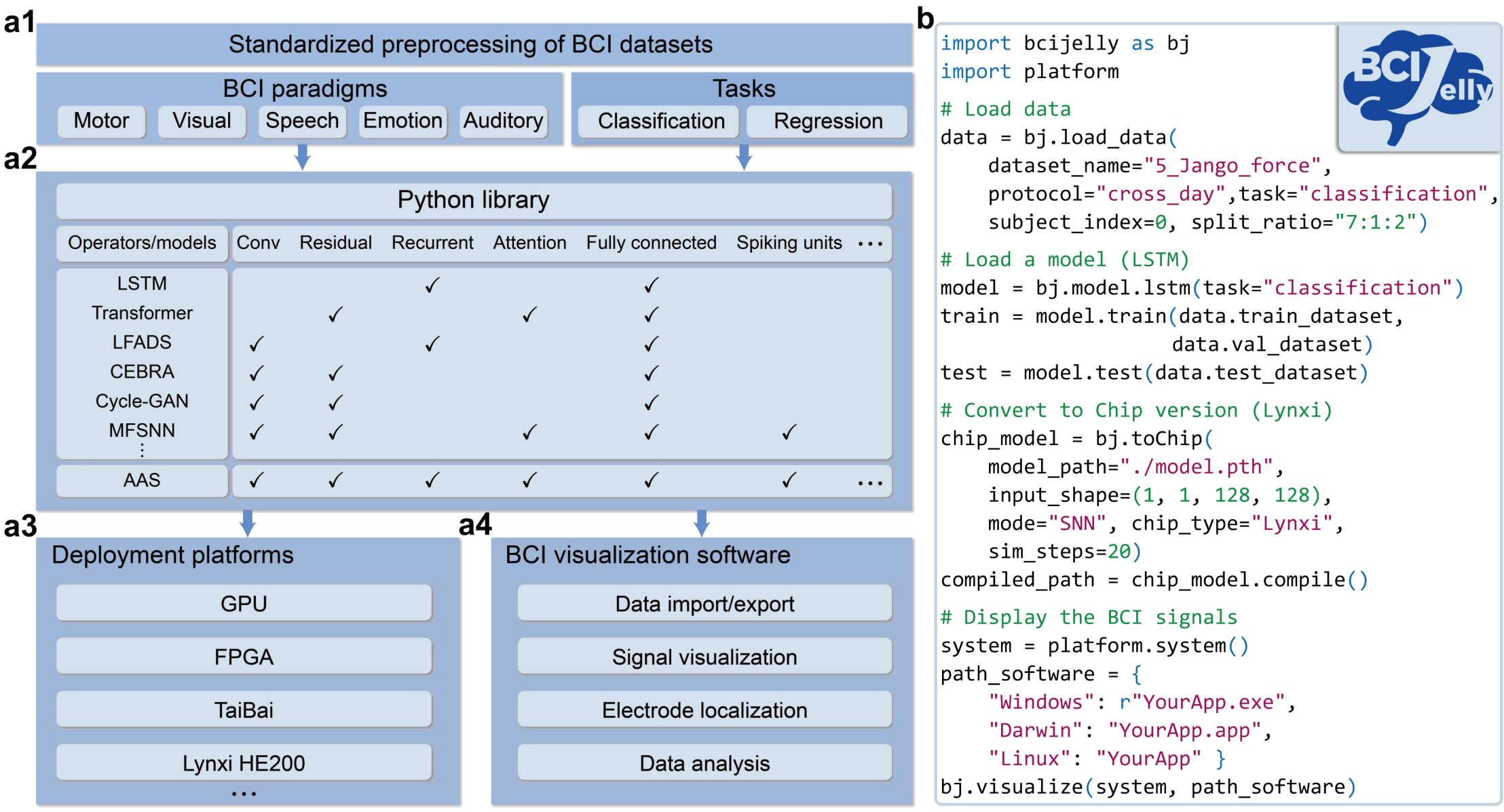

```python
import bcijelly as bj
import platform
# Load data
data = bj.load_data(
    dataset_name="5_Jango_force",
    protocol="cross_day",task="classification",
    subject_index=0, split_ratio="7:1:2")
# Load a model (LSTM)
model = bj.model.lstm(task="classification")
train = model.train(data.train_dataset,
                    data.val_dataset)
test = model.test(data.test_dataset)
# Convert to Chip version (Lynxi)
chip_model = bj.toChip(
    model_path="./model.pth",
    input_shape=(1, 1, 128, 128),
    mode="SNN", chip_type="Lynxi",
    sim_steps=20)
compiled_path = chip_model.compile()
# Display the BCI signals
system = platform.system()
path_software = {
    "Windows": r"YourApp.exe",
    "Darwin": "YourApp.app",
    "Linux": "YourApp" }
bj.visualize(system, path_software)
```

**Fig. 1 | Overview of the BCIJelly computational ecosystem. a,** Schematic of the BCIJelly framework integrating data preprocessing, model development, hardware deployment and visualization. **a1,** Standardized preprocessing of heterogeneous BCI datasets across five paradigms, including motor, visual, speech, emotion and auditory tasks, with support for both classification and regression analyses. **a2,** Modular model library organized at two levels. The algorithm level contains complete decoders that serve as ready-to-use baselines (LSTM, Transformer, LFADS, Cycle-GAN and

others). The module level contains reusable components (convolutional, residual, attention, fully connected and others) that can be composed into new architectures. AAS draws from the module pool to automatically sample, assemble and evaluate module combinations, generating task-specific decoders without manual architecture design. **a3,** Models trained on GPUs are compiled and deployed onto neuromorphic chips, including TaiBai and Lynxi HE200. FPGAs serve as presilicon prototypes for chips still in fabrication. **a4,** Interactive visualization software providing a single graphical interface for data import and export, signal visualization, electrode localization and data analysis. **b**, Representative workflow using the *bcijelly* Python library, in which loading a dataset, training a decoder, converting it for neuromorphic deployment and visualizing the signals are expressed in just a few lines of code.

## Preprocessing of standardized data across paradigms and species

BCIJelly supports five widely studied BCI paradigms, including motor, visual, speech, emotion and auditory tasks, with datasets recorded from humans, macaques and mice (**Fig. 2a**). To transform these heterogeneous recordings into a common representation, we implemented a standardized pipeline that converts continuous voltage traces into multichannel spike trains (**Fig. 2b**). Each single-channel signal is first bandpass filtered between 250 and 5,000 Hz, after which spikes are detected by a per-channel adaptive procedure that retains events falling between two amplitude bounds ($\mu \pm 4\sigma$ and $\mu \pm 8\sigma$, **Methods**). The spikes detected from all the channels are then assembled into a multichannel spike train.

Because trial duration periods vary across recordings but neural networks typically require fixed-length inputs, BCIJelly provides two binning strategies tailored to its two supported task types (**Fig. 2c**). For classification, where an entire trial maps to a single label, each trial is divided into a fixed number of steps, with the bin size adapted per trial so that trials of different duration periods are compressed into the same number of steps. For regression, where the neural signal is decoded point by point into a continuous behavioral trajectory, BCIJelly instead uses a uniform bin size shared across all trials, typically between 20 and 50 ms (**Supplementary Table 1**).

We next asked whether this binning step preserves the information contained in the underlying spike trains. On four representative datasets, two classification (center-out reaching) and two regression (reaching and random target tracking) tasks, we first compared the total spike count of each trial before and after binning. The two quantities were perfectly correlated across all the datasets (Pearson $r = 1.00$, **Fig. 2d, top**), confirming that no spikes were lost during binning. We next tested whether the temporal and geometric structures of the neural representations were also preserved. For each dataset, we computed pairwise distances between 300 trials in two spaces, the van Rossum distance in the original spike-train space and the Euclidean distance in the binned space, computed on the flattened time-by-channel representation. The two distance structures were highly consistent across datasets (Spearman $r = 0.94$ to 0.99, Mantel test, $p < 0.001$, **Fig. 2d, bottom**), indicating that the binned representation faithfully retains the temporal and geometric structure of the original neural activity while providing a compact input format for downstream models.

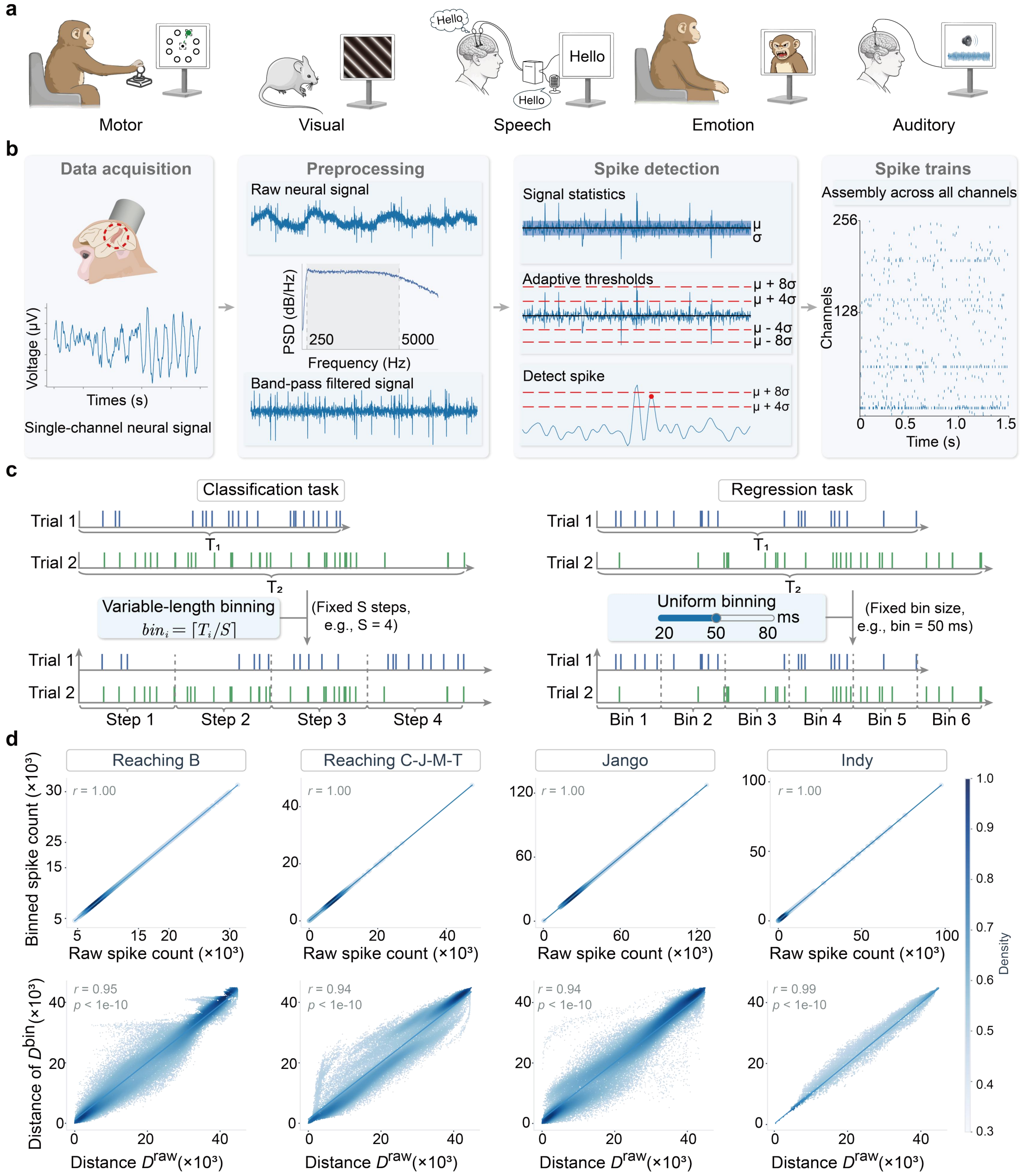

**Fig. 2 | Standardized preprocessing of heterogeneous BCI datasets. a**, BCIJelly supports five widely studied BCI paradigms (motor, visual, speech, emotion and auditory), with datasets recorded from humans, macaques and mice. **b**, Pipeline for converting raw recordings into multichannel spike trains. Each single-channel signal is first bandpass filtered between 250 and 5,000 Hz. Spikes are then detected by a per-channel adaptive threshold procedure that retains events whose amplitudes fall between $\mu \pm 4\sigma$ and $\mu \pm 8\sigma$, where $\mu$ and $\sigma$ are the mean and standard deviation of the filtered signal, respectively. The detected spikes from all the channels are finally assembled into a multichannel spike train. **c**, Two binning strategies for handling

variable trial duration periods. Left, variable-length binning for classification tasks, where each trial is compressed into a fixed number of steps S with the bin size set per trial as labeled. Right, uniform binning for regression tasks, where all trials share a fixed bin size (typically between 20 and 50 ms). **d**, Validation of the binning procedure on four representative datasets (two classifications, Reaching B and Reaching C-J-M-T, center-out reaching tasks; two regressions, Jango and Indy, reaching and random target tracking tasks). Top, per-trial total spike counts before and after binning are perfectly correlated (Pearson $r$ = 1.00). Bottom, pairwise distances between 300 trials in the raw spike-train space (van Rossum distance, $D^{bin}$) and in the binned space (Euclidean distance on the flattened time-by-channel representation, $D^{raw}$) are highly consistent (Spearman $r$ = 0.94 to 0.99, Mantel test, $p < 0.001$). The color intensity indicates the point density.

## A modular model library and automated architecture search

A central feature of BCIJelly is its 80-module library, which is grouped into four functional categories (fully connected, backbones, convolutional and attention; **Fig. 3a, left**). These modules form the basis of the decoding algorithms collected in BCIJelly, providing a common set of modular components for architectures ranging from classical to recent state-of-the-art decoders (**Fig. 3a, right**). The same modules can be freely recombined to generate architectures beyond those included in the existing collection.

Building on this library, a panel of representative decoders (LSTM, Transformer, LFADS, Cycle-GAN and Stabilization) was implemented within BCIJelly and evaluated across all five paradigms. Each decoder was tested under two settings, single-day decoding (training and testing on the same recording day) and cross-day decoding (testing on days held out from training; **Fig. 3b, Supplementary Table 2,** and **Supplementary File 3**). No decoder dominated across the paradigms or across settings, with the best-performing architecture changing from one dataset to the next (**Supplementary Tables 3-4**). This confirms that no single hand-designed model is universally optimal, motivating an automated approach to architecture selection.

We therefore developed automated architecture search (AAS), implemented in BCIJelly as the *SearchAgent* component, to search the module library for high-performing combinations on a given dataset (**Fig. 3c** and **Supplementary Table 12**). For each dataset, AAS iteratively samples candidate architectures from the library, trains them on the training set and evaluates them on a validation set. Candidates whose validation metric exceeds a predefined threshold are retained, and the rest are discarded. The loop stops when $K$ valid candidates have been collected or when a maximum number of search rounds is reached (**Supplementary Table 10**). AAS then performs an ensemble selection step over the retained candidates to produce a single combined model, whose performance is reported on a held-out test set (see Methods).

To assess how AAS performs in practice, we ran five independent searches on representative classification and regression datasets under the cross-day decoding setting (**Fig. 3d**). Despite limiting each search to no more than six candidate evaluations (see Methods), the five runs explored very different module combinations (**Fig. 3d**, overlap matrices) yet consistently produced strong decoders. The best run reached a median accuracy of 0.71 on the classification task and a median $R^2$ of 0.80 on the regression task, evaluated across 11 and 13 held-out test days, spanning 35 and 88 days respectively. These results show that a small number of AAS runs can identify competitive starting models, reducing the need to design architectures from scratch.

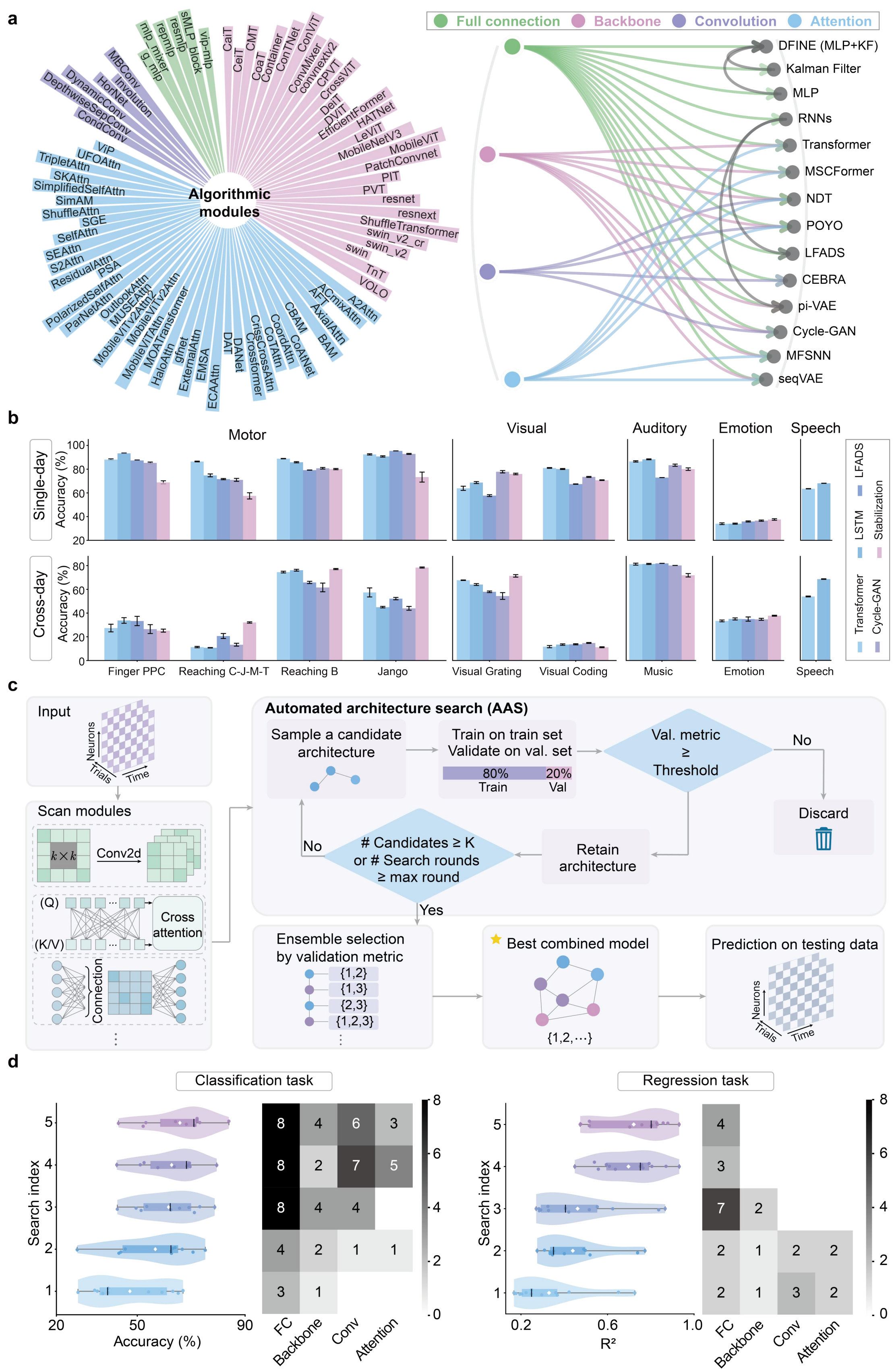

**Fig. 3 | Modular decoders and automated architecture search in BCIJelly. a,** The algorithmic module library in BCIJelly. Left, the 80 modules currently included in

BCIJelly, grouped into four functional categories (fully connected, backbones, convolutional and attention). Right, schematic showing how the same modules are combined to construct a range of established and recent decoding architectures collected in BCIJelly. **b**, Performance of five representative decoders (LSTM, Transformer, LFADS, Cycle-GAN and Stabilization) implemented in BCIJelly and benchmarked across all five paradigms (motor, visual, auditory, emotion and speech). Top, single-day decoding accuracy. Bottom, cross-day decoding accuracy. The bars indicate the mean across test days, and the error bars indicate one standard deviation. No decoder is dominant across the paradigms or across the two evaluation settings. **c**, Pipeline of AAS. Starting from an input neural recording, AAS scans the module library and assembles candidate architectures from reusable modules. Each candidate is trained on 80% of the data and validated on the remaining 20%, and only candidates that meet the validation threshold are retained. The retained candidates are passed to an ensemble selection step, which combines them into the best combined model for evaluation on the held-out test set. **d**, Performance of AAS across five independent searches on a representative classification dataset (left panel) and a representative regression dataset (right panel), under the cross-day decoding setting. Each violin shows the distribution of test-day performance for one search, with markers indicating individual test days. Adjacent matrices report the number of overlapping modules between pairs of searches, showing that different runs converge to distinct module combinations. The best run reached a median accuracy of 0.71 on the classification task (11 test days spanning 35 days) and a median $R^2$ of 0.80 on the regression task (13 test days spanning 88 days).

**Closed-loop AAS with large language models**

The architecture search (**Fig. 3**) described thus far was designed for a single task at a time, and the module library was explored by random sampling. To handle multitask decoding and to replace random sampling with a guided design, we extended AAS with an LLM that plans, diagnoses and refines decoders in a closed loop (**Fig. 4a** and **Supplementary Video 2**). The agent reads three structured files that define the search context, a task specification (Task.md, covering the training, validation and test split, the species and the task type, including single-task and multitask decoding), a description of every available module (Module.md) and a running history of previously searched architectures with their parameter settings and validation and test scores (History.json). Guided by this context, the LLM first plans the search and pretrains a shared encoder on the training data, which is then frozen. It then enters a decoder search and selection stage, in which candidate decoders are assembled according to the plan and evaluated on the validation set. A decoder whose validation metric exceeds the threshold is subjected to hold-out testing, and its architecture and scores are written back to the history file in a knowledge evolution step. A decoder that falls below the threshold instead enters a diagnosis and design loop, in which the LLM examines why validation failed, designs a new module, checks that its shape is compatible with the data dimensions and returns it to the assembly stage for re-evaluation. The loop terminates under explicit stopping conditions, after which the accumulated knowledge is committed to the history file.

We evaluated this LLM–driven search on five invasive BCI datasets (Reaching B, Visual Coding, Jango, Indy and Finger PPC) that together span classification and regression tasks and recordings from humans, macaques and mice. The shared encoder (**Extended Data Fig. 5** and **Supplementary Table 20**) was designed using Claude Opus-4.7 and reused as a common backbone, while the decoder search was run independently under three LLM backends (DeepSeek-V4 Pro, ChatGPT-5.5 and Claude Opus-4.7), each of which planned and searched its own decoder. We compared these searched decoders against three current state-of-the-art pretrained models (NDT1, NDT2 and POYO) and against our own pretrained model (UniBCI[25], **Extended Data Fig. 7**), evaluating each on the single-task setting (**Extended Data Fig. 6**), on the classification and regression tasks, and on the cross-species joint-training settings (human and macaque, human and mouse, macaque and mouse, and all three species together; **Fig. 4b** and **Supplementary File 2**). Statistical significance was assessed by paired *t* tests (**Supplementary File 1**).

Across these settings, the encoder–decoder combinations produced by LLM–driven AAS were competitive with state-of-the-art pretrained models and exceeded them in selected cross-species settings (**Fig. 4b**). The greatest differences appeared in the cross-species joint-training settings, where on the Jango and Reaching B datasets the searched decoders scored significantly higher than the best pretrained baseline (paired t-test, $P < 0.05$). On the remaining datasets, and on the classification and regression tasks evaluated separately, the searched decoders and the best baselines did not differ significantly. On the Indy dataset, NDT1 scored significantly higher than the searched decoders under the macaque and mouse and the all-species settings. The three LLM backends did not differ significantly from one another.

UniBCI performed competitively on the Finger PPC and Visual Coding datasets, matching the best pretrained baselines without a significant difference and achieving the highest numerical scores on Finger PPC in several settings. On the other datasets it scored below the state-of-the-art baselines.

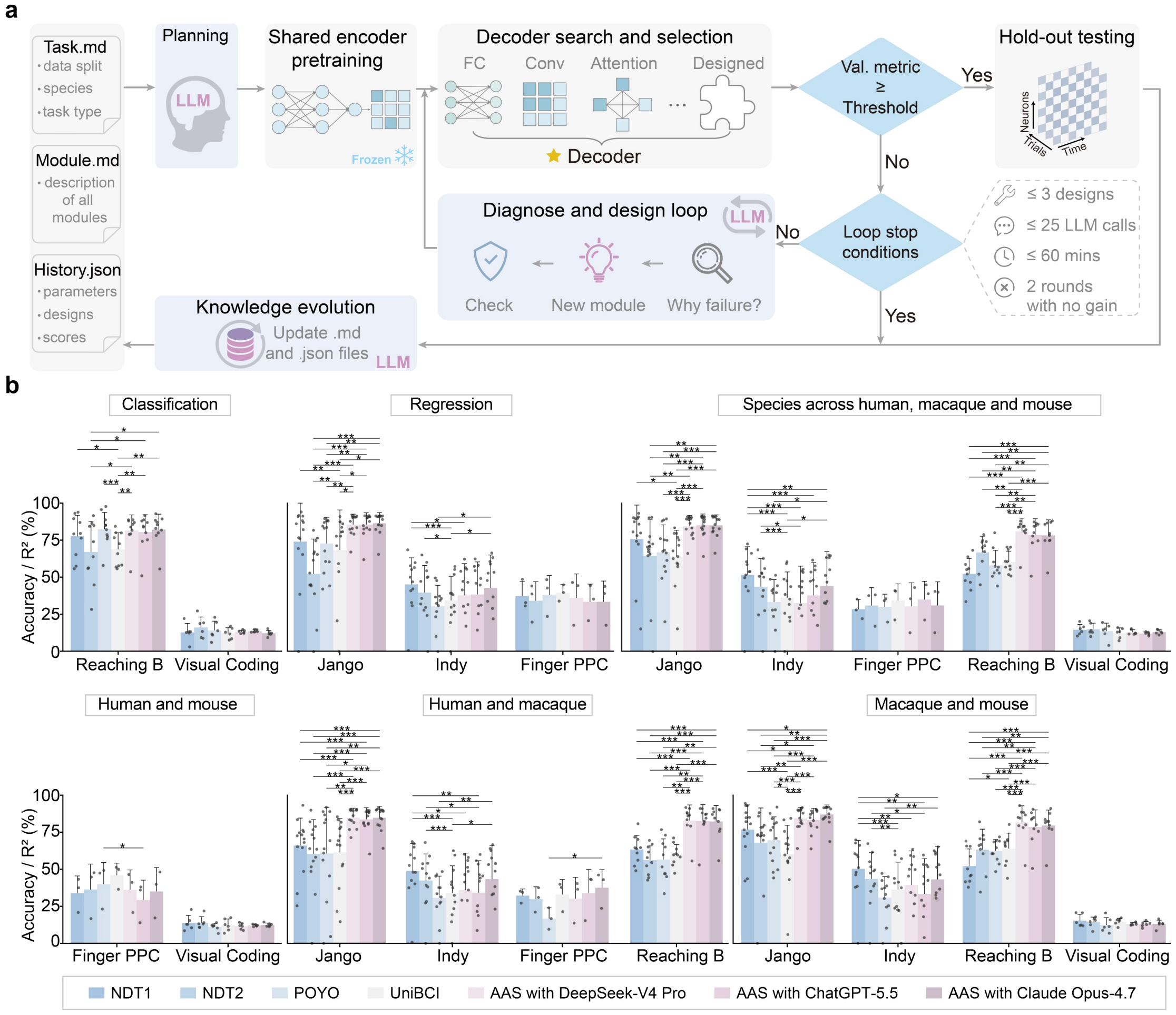

**Fig. 4 | Closed-loop automated architecture search with large language models. a,** The LLM–driven AAS pipeline. The LLM reads three structured files (Task.md, Module.md and History.json), pretrains a shared encoder that is then frozen, and assembles and evaluates the candidate decoders. Decoders meeting the validation threshold proceed to hold-out testing and are recorded in a knowledge evolution step, while those below it enters a diagnosis and design loop that proposes new modules for re-evaluation. Stopping conditions are shown on the right (at most 3 new module designs, 25 LLM calls, 60 minutes, or 2 consecutive rounds with no gain). **b,** Decoding performance across five invasive BCI datasets (Reaching B, Visual Coding, Jango, Indy and Finger PPC) spanning classification and regression tasks and recordings from humans, macaques and mice. Searched decoders under three LLM backends (DeepSeek-V4 Pro, ChatGPT-5.5 and Claude Opus-4.7) are compared with three state-of-the-art pretrained models (NDT1, NDT2 and POYO) and with our pretrained model (UniBCI[25]). Panels show the classification and regression tasks and

the cross-species joint-training settings (human and macaque, human and mouse, macaque and mouse, and all three species together). Accuracy is reported for classification and the coefficient of determination ($R^2$, as a percentage) for regression. The bars indicate the mean across recording days (or sessions, for datasets recorded in a single session), the dots show individual recording days, and error bars indicate one standard deviation. Statistical significance was assessed using a paired t-test (see n pairs, *P* values and *t* values in **Supplementary File 1**). *$P < 0.05$, **$P < 0.01$, ***$P < 0.001$.

## Hardware-aware deployment across heterogeneous platforms

Once a decoder is trained, deploying it onto a new chip typically requires substantial engineering because each piece of hardware supports a different set of native operators and instructions. BCIJelly addresses this with a unified deployment pipeline that compiles trained models for any of the supported chips (**Fig. 5a**). For standard artificial neural networks (ANNs), the trained model is parsed directly into a computational graph that the pipeline then compiles for the chosen platform. For spiking neural networks (SNNs), an additional node conversion step first translates PyTorch operators into their spike-based equivalents, after which the converted graph follows the same parsing and compilation steps as the ANNs do. In practice, BCIJelly exposes the choice of ANN or SNN deployment as a single argument to its conversion API (**Fig. 1b**).

To understand which models can be efficiently mapped onto each platform, we cataloged the native operator coverage of three representative chips (**Fig. 5b**). These included a general-purpose GPU and two neuromorphic chips, Lynxi and TaiBai[21]. The GPU offered the broadest operator coverage. For the ANNs, the building modules that the BCIJelly decoders rely on, spanning convolution, batch normalization, fully connected layers, recurrent units and attention, were natively supported on both the GPU and the Lynxi toolchain, so that the standard ANN decoders were compiled to either platform without architectural modification. AAS–generated ANN architectures produced were assembled entirely from these natively supported operators and could therefore be deployed directly (**Extended Data Fig. 4**c). The two neuromorphic chips traded some of this generality for the native support of spike-based computation, including spike-driven inference and sparse weight encoding. As a result, the SNN decoders in BCIJelly ran not only on the GPU but also on both neuromorphic platforms, making Lynxi and TaiBai well suited as deployment targets for the spiking models in BCIJelly.

We next evaluated deployment on three platforms using the multiscale fusion enhanced spiking neural network (MFSNN[26]) algorithm, a spiking model previously developed for motor BCI decoding (**Fig. 5c**). All three platforms produced essentially identical decoding performance, confirming that operator conversion preserves model performance during deployment. Compared with 16.48 W on the GPU, the two neuromorphic chips consumed only a fraction of this power (Lynxi 0.57 W and TaiBai 0.34 W), corresponding to approximately 30-fold and 50-fold reductions, respectively. These results demonstrate that dedicated neuromorphic hardware can substantially reduce the energy cost of deploying spiking decoders.

A key contributor to this large power reduction is the operator conversion and graph-level optimization that BCIJelly performs during compilation, which we illustrate on the MFSNN (**Fig. 5d**). At the operator level, BCIJelly fuses each sequence of Conv2D, BN2D and leaky integrate-and-fire (LIF) layers into a single spike-conv2D operator that the target chip can execute natively. At the graph level, BCIJelly further optimizes the placement of the resulting computational graph onto the chip's core array of the chip, reducing the number of cores needed to host the MFSNN from 113 to 33 (a more than threefold reduction) without loss of decoding accuracy (**Fig. 5c**).

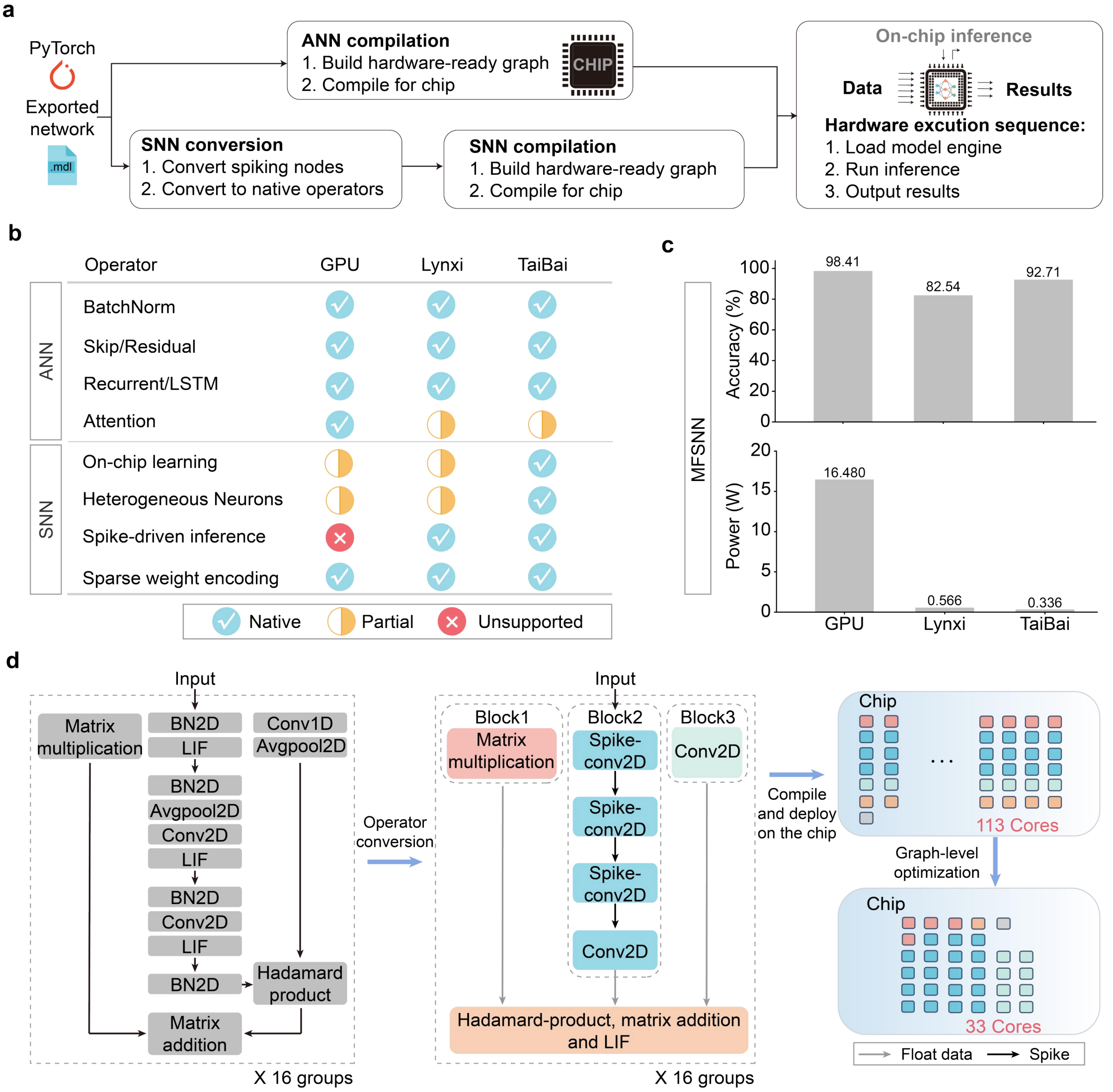

**Fig. 5 | Compilation and deployment of decoders in BCIJelly. a**, Unified deployment pipeline. PyTorch-trained ANN models are parsed into a hardware-ready computational graph that is then compiled for the target chip (**Extended Data Fig. 4**). SNN models pass through an additional node conversion step that translates PyTorch operators into spike-based equivalents, after which the converted graph follows the same parsing and compilation steps as ANNs. The choice of ANN or SNN deployment is set by a single argument to the conversion API. **b**, Native operator coverage of three representative chips, a general-purpose GPU, the Lynxi neuromorphic chip and TaiBai, an in-house neuromorphic chip. Fully supported (green check), partially supported (half-orange) and not supported (red cross). **c,** Deployment evaluation of the MFSNN on the three platforms. **Left**, decoding accuracy is essentially identical across the GPU, Lynxi and TaiBai, confirming that

operator conversion preserves model performance during deployment. **Right**, power consumption is reduced by approximately 30-fold on Lynxi (0.57 W) and 50-fold on TaiBai (0.34 W) compared with the GPU (16.48 W, NVIDIA RTX 4060). **d**, Operator conversion and graph-level optimization of the MFSNN during compilation. **Left**, the original MFSNN architecture, in which long sequences of Conv2D, BN2D and LIF layers are interleaved with matrix multiplication, the Hadamard product and matrix addition operations. **Middle**, after operator-level optimization, each Conv2D-BN2D-LIF sequence is fused into a single spike-conv2D operator that the target chip can execute natively. **Right**, after graph-level optimization, the computational graph is placed onto the core array of TaiBai with a reduction in the core count from 113 to 33.

## A graphical interface for the BCIJelly workflow

To make BCIJelly accessible beyond command-line use, we developed a companion visualization software that combines neural signal inspection, spike detection, electrode localization, and decoder configuration and output visualization in a single graphical interface. The software supports recordings from humans, macaques and mice, and covers the full range of paradigms supported by BCIJelly (**Fig. 6a**). For each species, the recording brain region is highlighted on a species-specific 3D brain and shown together with the corresponding behavioral paradigm, with an auditory (speech) task in the right superior temporal gyrus for the human, an eight-direction reaching task in the primary motor cortex for the macaque, and a visual grating task in the primary visual cortex for the mouse. This presentation maps recording sites to brain regions consistently across human, macaque and mouse experiments.

Its main interface combines several synchronized panels (**Fig. 6b**). Users select recording channels from a list, view the recording site on a species-specific 3D brain, watch the behavioral video stream from the task, and inspect multichannel local field potential (LFP) traces alongside recording statistics such as the sampling rate and total duration. The full BCIJelly spike detection pipeline is also accessible through interactive controls, with adjustable bandpass cutoffs and threshold levels, and detected spikes are overlaid on the LFP traces for immediate visual feedback (**Fig. 6c**).

The software also serves as a front end for the BCIJelly decoder library. Any decoder in the library can be loaded and run on a chosen task (**Fig. 6d**). Decoding outputs are displayed side by side with the ground-truth behavior, allowing the decoder predictions to be compared directly with the actual behavior. This comparison is shown both for tasks that involve selecting one of several reach directions (classification; **Fig. 6e**) and for tasks that reconstruct a continuous trajectory (regression; **Fig. 6f**). The full workflow, from raw signals to decoded behavior, can thus be explored interactively without programming.

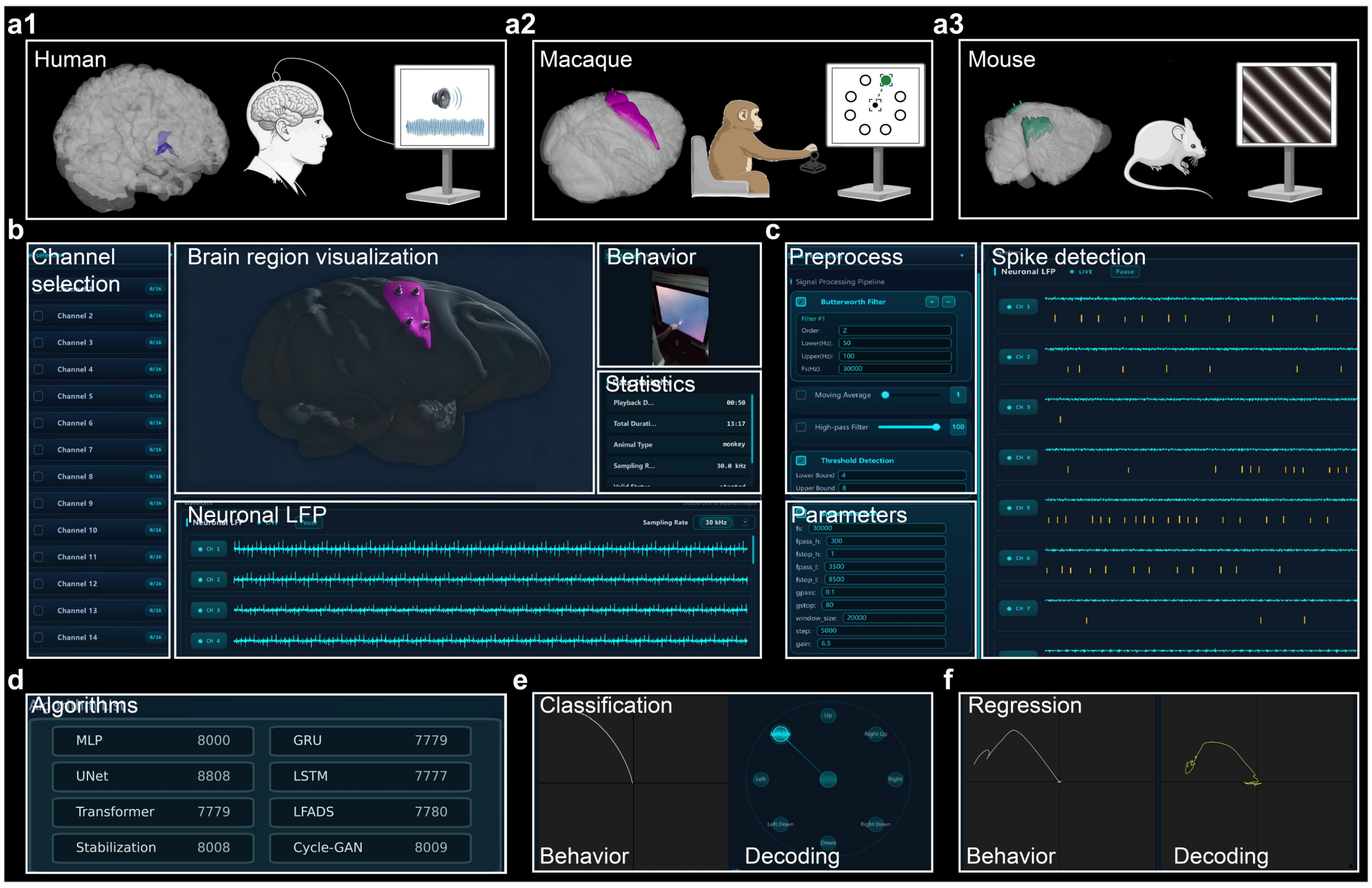

**Fig. 6 | Interactive visualization of neural signals and decoding outputs. a,** BCIJelly visualization software supports recordings from humans (a1), macaques (a2) and mice (a3), covering the full range of paradigms supported by BCIJelly. For each species, the recorded brain region (highlighted on a 3D brain, left) is shown alongside the experimental subject and the corresponding behavioral paradigm (right), namely an auditory (speech) task in the right superior temporal gyrus for the human, an eight-direction reaching task in the primary motor cortex for the macaque, and a visual

grating task in the primary visual cortex for the mouse. **b,** Main interface of the software. Users select recording channels from a list (left), view the recording site on a species-specific 3D brain (center), watch the behavioral video stream from the task (top right), and inspect multichannel local field potential (LFP) traces (bottom) alongside recording statistics including the sampling rate and total duration (right). **c,** Interactive spike detection. Bandpass cutoffs and threshold levels are set in the Preprocess and Parameters panels (left), and detected spikes are overlaid on the LFP traces (right) for immediate visual feedback. **d,** Decoder configuration panel. Any decoder in the BCIJelly library, including the MLP, GRU, UNet, LSTM, Transformer, LFADS, Stabilization and Cycle-GAN, can be loaded through the algorithm list and assigned to a chosen task. **e,** Classification task. The actual reach direction (Behavior, left) is shown alongside the predicted reach direction selected from eight discrete targets (Decoding, right). **f,** Regression task. The actual reach trajectory (Behavior, left) is shown alongside the reconstructed continuous trajectory (Decoding, right).

## Discussion

BCIJelly is a unified computational ecosystem newly developed for BCI research that integrates preprocessing, benchmark decoders, automated architecture search and hardware-aware deployment within a single Python framework. A unique property of BCIJelly is that it spans the full BCI workflow, from raw neural recordings through decoder development to deployment on neuromorphic chips, with a modular API that allows components to be combined, swapped or extended on demand. Drawing from a library of 80 reusable modules, AAS builds task-specific decoders without manual architecture design, and its LLM–driven extension supports guided single-task, multi-task and cross-species decoder design. In parallel, compared to general-purpose GPU, the *toChip* pipeline compiles trained models for neuromorphic hardware with substantial reductions in power consumption (**Extended Data Figs. 3–4**). An interactive visualization platform makes these capabilities accessible through a graphical interface, enabling exploration of the full workflow without programming.

We validated BCIJelly across five BCI paradigms (motor, visual, speech, emotion and auditory) and recordings from humans, macaques and mice, supporting its use as an extensible infrastructure for diverse BCI workflows.

BCIJelly integrates capabilities that are typically developed and used separately in BCI research. Recent suites such as NeuroAI and TorchBrain consolidate data handling, training and benchmarking for neural data more broadly, whereas BCIJelly is built around the decoding-to-deployment pipeline specific to BCI research, and two of its design choices follow directly from the demands of this setting. First, because decoder performance varies substantially across datasets and paradigms (**Extended Data Fig. 1, Supplementary Tables 5-9** and **Supplementary File 4**), BCIJelly couples a library of reusable computational modules with an automated search procedure that assembles task-specific architectures without manual model design, rather than committing to a fixed decoder. The LLM–driven extension of AAS generalizes the search process beyond random sampling from a fixed module library. It retains the ability to construct single-task decoders (**Extended Data Fig. 6**) and extends the same framework to guided multi-task and cross-species decoder design. Its similar performance across the three tested LLM backends suggests that this workflow is not tied to a single language model implementation (**Fig. 4b** and **Supplementary File 2**). Compared with recent LLM–driven systems for algorithm design ($A_2$DEPT[27]) and AI architecture discovery (ASI-Arch[28]), LLM–driven AAS is specialized for constructing a BCI decoder. Within BCIJelly, this search procedure operates alongside standardized neural data preprocessing, decoder evaluation and visualization, rather than as a standalone architecture-discovery system. The comparable performance of the searched encoder-decoder models against pretrained baselines further suggests that guided task-specific design can be competitive with general pretraining in this setting. Second, because many practical BCI applications require low-power operations, trained decoders are carried through to neuromorphic hardware via the *toChip* pipeline, rather than stopping at GPU training. Together, these capabilities establish a continuous workflow from raw neural recordings to hardware execution, reducing the engineering effort required to evaluate and deploy new decoding approaches across diverse BCI settings.

Several limitations of the current framework should be noted. BCIJelly currently focuses on invasive neural recordings, including intracortical spike trains and electrocorticography. Noninvasive modalities, including scalp EEG and functional near-infrared spectroscopy, are central to many BCI applications and are addressed by complementary frameworks such as MetaBCI[29]. Extending BCIJelly to these modalities is therefore an important next step. The decoder library covers the major methodological families used in BCI research, but only one spiking neural network decoder (MFSNN) is currently included. Expanding this decoder class would be valuable because spiking models are well matched to intracortical recordings, where the underlying signal is itself a population of spikes. The *toChip* pipeline currently targets two neuromorphic platforms (Lynxi HE200 and TaiBai). All the TaiBai results reported here were obtained using its FPGA prototype, and the corresponding ASIC implementation has been taped out and is currently in fabrication. Some decoders produced by AAS can be deployed on neuromorphic chips (**Extended Data Figs. 3-4**), but the LLM–driven encoder-decoder models were not deployed because the shared encoder contains Transformer blocks that are not yet supported by the current chip toolchains. The LLM–driven search was run once per task under a fixed pipeline seed, so uncertainty reflects held-out recording days rather than language-model stochasticity. The LLM–driven comparison also isolates the decoder search by using a shared encoder architecture designed with Claude Opus-4.7, and therefore does not test whether each LLM could independently design the full encoder-decoder architecture. Finally, the visualization platform currently runs on Windows only, with Linux and macOS versions under active development.

Collectively, by unifying the full BCI workflow into a single computational ecosystem, BCIJelly both lowers the barrier to constructing task-specific decoders through automated architecture search and bridges the gap between model training and execution on neuromorphic chips. We anticipate that this integration will support the translation of advanced decoding methods to energy-efficient BCI systems for neural interface applications.

## Methods

### Datasets and standardized preprocessing.

**Motor BCI datasets.** We used 13 motor BCI datasets covering four task categories. For center-out reaching, we used Reaching C-J-M-T[30], Jango[13], Reaching C-M[13] and Link CO[31]. For random target tracking, we used Link RTT[31], Indy[32] and FALCON M2[33]. The remaining five datasets include FALCON M1-A[34] and Grasp P-G[13] for reach-to-grasp, Maze N-J[35] for delayed reaching with maze barrier, Finger RL[36] for finger movement and Finger PPC[37] for finger flexion. We additionally included an in-house dataset, Reaching B, in which a crab-eating macaque performed an eight-direction center-out reaching task while neural activity was chronically recorded from the primary motor cortex (M1). The trials with various time lengths were normalized to a fixed length of 50 time steps.

**Visual BCI datasets.** We used two visual BCI datasets, the Visual Coding[1] and the Visual Grating[38] datasets, spanning mouse and macaque recordings under multiple visual stimulus paradigms.

**Speech, auditory and emotion BCI datasets**. We used one dataset for each of these three paradigms. In humans, we used intracortical recordings during sentence production (Speech[3]) and cortical recordings during passive music listening (Music[5]). In a head-fixed macaque, we used ECoG recordings during passive movie viewing (Emotion[39,40]). Detailed information for each dataset, including species, recording protocol, citation and code examples, is available on the BCIJelly companion website (https://bcihub.github.io/BCIJelly-site).

**Spike detection.** For each recording, raw voltage traces were bandpass filtered with a Butterworth filter applied in zero-phase mode. Spikes were then detected on each channel independently using a sliding window procedure with two amplitude bounds. The filtered trace was divided into nonoverlapping windows, and within each window the mean $\mu$ and standard deviation $\sigma$ were computed. A spike was registered whenever the signal amplitude fell between the two bounds $\mu \pm \alpha\sigma$ and $\mu \pm \beta\sigma$, that is, whenever $\alpha\sigma < |V(t) - \mu| < \beta\sigma$, where $V(t)$ is the filtered voltage at time t. The lower bound ($\alpha\sigma$) excludes background noise, while the upper bound ($\beta\sigma$) excludes large transient artifacts. A refractory period was enforced to prevent redetection of the same spike. Filter cutoffs, filter order, window size, multipliers ($\alpha$, $\beta$) and refractory period were configured per dataset. As an example, for the GY_data dataset we used cutoff frequencies of 250 Hz and 5,000 Hz, a 5th-order Butterworth filter, a sliding window of 200,000 time points (approximately 6.67 s), $\alpha = 4$ and $\beta = 8$, and a refractory period of 34 time points (approximately 1.13 ms).

**Binning in BCIJelly.** For classification tasks, where an entire trial is mapped to a single label, each trial was divided into a fixed number of steps $S$. The bin size for trial $i$ was set to $bin^i = \lceil T_i / S \rceil$, where $T_i$ is the number of time points in trial $i$ and $\lceil \cdot \rceil$ denotes the ceiling function. This adaptive bin size yields the same output shape (channels × S) for all trials regardless of their original duration. For regression tasks, where the neural signal is decoded point by point into a continuous behavioral trajectory, a uniform bin size $b$ was used across all trials, typically between 20 and 80 ms. The number of bins per trial $S^i = \lceil T_i / b \rceil$ thus varies with trial duration. In both strategies, spikes within each bin were counted to produce the binned spike train. When the trial duration was not a multiple of the bin size, the final bin was zero-padded to the bin size. The step count $S$ (classification) and bin size $b$ (regression) were configured per dataset (**Supplementary Table 1**). This binning procedure was validated by comparing per-trial spike counts and pairwise spike-train distances (van Rossum vs Euclidean) between 300 randomly selected trials per dataset, with significance assessed by the Mantel test with 999 permutations (**Fig. 2d**).

## BCIJelly modules and models

**Module library.** BCIJelly provides a library of 80 reusable computational modules organized into four functional categories. These include 39 attention modules, 29 backbone modules, 6 convolutional modules and 6 fully connected modules. Attention modules implement attention-based mechanisms for context-dependent feature weighting (e.g., self-attention, cross-attention, and residual attention). Backbone modules encapsulate established neural network architectures used as high-level building units (e.g., ResNet, Swin Transformer, and DViT). Convolutional modules implement 1D and 2D convolution operations for spatial and temporal filtering (e.g., dynamic convolution, CondConv, and MBConv). Fully connected modules implement projection layers based on linear and multilayer perceptron architectures (e.g., RepMLP, gMLP, and MLP-Mixer). A complete list of modules, their parameter specifications and corresponding references is provided on the BCIJelly companion website (https://bcihub.github.io/BCIJelly-site/).

**Automated architecture search.** AAS is implemented in BCIJelly as the *SearchAgent* component. AAS automates the discovery of high-performing decoder architectures from the BCIJelly module library. Given a training set, a validation set and a held-out test set, AAS iteratively samples and evaluates the candidate decoder architectures, retains those that exceed a validation performance threshold, and combines the top candidates into a final ensemble decoder.

The candidate architectures are sampled uniformly at random from the four module categories of the BCIJelly library. Each candidate is a sequential composition of up to $D$ modules, subject to three diversity constraints. First, at most two modules from the same category are included per candidate. Second, modules from the backbone category are not repeated within a candidate. Third, modules from at least $N_min$ distinct categories must appear. Nonbackbone modules may appear multiple times subject to the first constraint.

The search proceeds for $R$ rounds. In each round, $M$ candidate architectures are sampled and each is independently trained on the training set and evaluated on the validation set. Candidates whose validation performance (accuracy for classification, $R^2$ for regression) exceeds the predefined threshold $\tau_val$ are retained. After all $R$ rounds, the $K$ best-performing retained candidates across all rounds are combined into the final ensemble by averaging their predictions on the held-out test set. For the Reaching B dataset (classification), we used $R = 3$ search rounds, $M = 4$ candidates per round, $D = 4$ maximum modules per candidate and $N_min = 2$ minimum distinct categories. For the Jango dataset (regression), we used $R = 3$, $M = 6$, $D = 3$ and $N_min = 1$. The validation threshold $\tau_val = 0.78$ and the ensemble size $K = 4$ were the same for both tasks. Five independent runs were performed per task (**Fig. 3d**), and the optimum architectures selected per task are listed (**Extended Data Fig. 2**).

**LLM–driven AAS.** All decoders were built on a single shared encoder (**Extended Data Fig. 5**) that mapped neural recordings to a common feature representation. The encoder was pretrained jointly across all the datasets by minimizing a multitask objective, which summed the cross-entropy loss over the classification datasets and the mean-square-error loss over the regression datasets through per-dataset linear heads. It was then frozen and reused across all the backends. The LLM–driven AAS drew on three context files, a task specification (Task.md), a description of the available modules (Module.md) and a history of evaluated architectures with their scores (History.json). At each round, the LLM returned a structured specification of a candidate decoder. This specification was resolved through the module-library registry into a trainable network at the encoder's hidden dimension, then trained on the training set and evaluated on the validation set using the splits defined above.

Candidates that exceeded a per-dataset validation threshold (**Supplementary Table 11**) were retained.

When a candidate fell below the threshold, the LLM designed a new module as an executable PyTorch definition. Before use, each module passed a safety and shape check, namely, a restricted-import check that permitted only torch, torch.nn, torch.nn.functional and math and rejected file; process or dynamic-evaluation calls; compilation in an isolated namespace; and a forward pass verifying the output shape and finite values. The module was then assembled into a decoder and evaluated, and was retained only if that decoder ran successfully and reached the threshold. The retained modules were written back into the library and described in Module.md for use in later searches. The search stopped at the first of four conditions, at most three new module designs, 25 LLM calls, 60 minutes, or two consecutive rounds without improvement. The single best-performing candidate was then selected as the final decoder. Each backend was accessed through its provider's API with a structured JSON response, and the encoder, thresholds and stopping conditions were identical across the backends. Because LLM outputs are stochastic, we fixed the random seed of the non–LLM pipeline (controlling data-loader shuffling, initialization and training), so that the results reproduce exactly given a fixed set of LLM outputs, and logged the full prompt and response transcript of every search for auditing and replay. The performance is reported as the distribution across held-out test days or sessions (**Fig. 4b**), which quantifies generalization across recording days rather than variability across random seeds.

As LLM–driven AAS proceeded, newly designed modules accumulated in the library and the record of evaluated architectures grew, thus, the context available to the LLM could in principle expand without bound. To keep the search scalable, we decoupled the stored library from the context shown to the LLM. The full library and history were kept on disk, but at each round the LLM saw only a fixed-size, performance-ranked slice of each module category. This slice combines the built-in modules with the most frequently selected modules, modules relevant to the current task (by provenance, prior performance and keyword overlap), and the most recently designed modules, each with a one-line summary. The context size therefore depended on the slice rather than on the library. Always exposing the highest-performing modules kept the search from regressing, while a few slots reserved for recent designs allowed new modules to be tried and promoted. Lower-performing, older modules were not shown to the LLM but remained in the library and were available to the search. With the context held at about 43 modules while the library grew beyond 100, the search remained tractable and its accuracy did not degrade, allowing the library continue expanding without retraining or re-tuning the search.

**Decoder library.** In addition to the module library, BCIJelly provides a collection of 15 complete decoder implementations covering the major methodological families used in BCI research. Linear and shallow models include the Kalman filter[41] and a multilayer perceptron[42] (**Supplementary Table 19**). Recurrent models are represented by LSTM[43]. Transformer-based models include a standard Transformer[44], MSCFormer[45], NDT1[9], NDT2[10] and POYO[11] (**Supplementary Tables 13-14** and **17**), which capture spatiotemporal structure through attention mechanism. Latent variable models include DFINE[46] and LFADS[12] (**Supplementary Table 15**), which infer the low-dimensional dynamic structure underlying neural activity. Cross-day alignment methods include seqVAE[47], Cycle-GAN[13], Stabilization[48] and NoMAD[14] (**Supplementary Table 16**), which align neural representations across recording days. Finally, the MFSNN[26] implements a multiscale fusion spiking neural network for deployment on neuromorphic hardware (**Supplementary Table 18**). All implemented decoders share a unified training protocol. Models were trained with the Adam optimizer. The classification models used cross-entropy loss, whereas the regression models used mean squared error. The per-dataset training parameters (number of epochs, learning rate, hidden dimensionality; **Supplementary Note 1**) are listed on the BCIJelly companion website (https://bcihub.github.io/BCIJelly-site/).

**Single-day and cross-day decoding.** Within each recording day, the first 80% of the trials were used for training and the remaining 20% were used for testing (single-day decoding). For cross-day decoding, the first 80% (or 70% for datasets with fewer recording days) of the recording days were used for training, and the remainder were used for testing. The day-by-day split details are provided (**Supplementary Tables 2 and 5**).

**BCIJelly API and example usage.** The Python implementation of BCIJelly is written in PyTorch and NumPy. It provides a unified application programming interface (API) that integrates data loading, model construction, training, evaluation, automated architecture search and neuromorphic chip deployment in a single framework. The library follows a three-step workflow of data loading, model construction, and training and evaluation.

Datasets are loaded through *bcijelly.load_data*, which returns a structured DataLoadResult containing in-memory training, validation and test arrays.

```
import bcijelly as bj
loaded = bj.load_data(
        dataset_name="5_Jango",
        protocol="single_day",
        task="regression",
        subject_index=0,
        day_index=0,
        split_ratio="7:1:2")
```

The protocol argument selects between single-day and cross-day loading paradigms, the task argument selects the label-key convention (classification, regression or speech), and the *split_ratio* controls the training, validation and test partitions. In single-day mode, trials within a recording day are split chronologically. In cross-day mode, the recording days themselves are partitioned into training, validation and test blocks. Custom datasets are supported through optional *x_keys* and

*y_keys* arguments that override the default feature- and label-key conventions.

Model instances are constructed through dynamic factories under *bcijelly.model*. Each factory accepts a task argument and returns a *BaseModel* adapter with unified *.train(...)* and *.test(...)* methods.

```
model = bj.model.transformer(task="regression")
train_result = model.train(
                 train_dataset=loaded.train_dataset,
                 val_dataset=loaded.val_dataset,
                epochs=12,
                device="auto")
test_result = model.test(test_dataset=loaded.test_dataset)
```

Fifteen decoder families are currently exposed through factories, namely *kf*, *mlp*, *transformer*, *lstm*, *lfads*, *cycle_gan*, *stabilization*, *fenet*, *dfine*, *ndt1*, *ndt2*, *poyo*, nomad, *seqvae* and *mfsnn*. The task-compatibility matrix between models and tasks is queried through *bcijelly.model.get_supported_tasks*. For cross-day evaluation, *model.test* can be invoked per recording day on *loaded.test_day_datasets* to produce day-by-day metrics.

AAS is implemented in the BCIJelly Python package as *bcijelly.SearchAgent*, an end-to-end search-and-ensemble API. Given training, validation and test datasets, this API samples candidate architectures from the BCIJelly module library, trains and evaluates each candidate on the validation set, retains top-performing candidates and selects an ensemble decoder for testing.

```
from bcijelly import SearchAgent
agent = SearchAgent(
            task_type="classification",
            max_search_rounds=3,
            max_models_per_round=4,
            retain_top_k=4,
            val_threshold=0.78,
```

```
            ensemble_method="mean")
    agent.fit(loaded.train_dataset, loaded.val_dataset)
    #  cross-day test input for SearchAgent
    by_day_test_data = dict(
                        ("day_" + str(i), ds)
                        for i, ds in enumerate(loaded.test_day_datasets or []))
    result = agent.test(by_day_test_data)
```

The *SearchAgent* API supports configurable architecture constraints, including the maximum depth of sampled architectures (*max_depth*), the maximum number of modules per type within a candidate (*max_per_type*), the minimum number of distinct module types per candidate (*min_distinct_types*) and intratype repetition rules (*no_repeat_within_types*, *allow_repeat*). Three ensemble strategies are available through the *ensemble_method*, namely mean averaging, stacking and an automatic mode that selects between them (**Supplementary Table 10**).

LLM–driven AAS is exposed through *bcijelly.LLMSearchAgent*, which wraps the closed-loop search procedure described above. The API uses a task specification, module descriptions and a search-history file to define the search context, and applies explicit stopping conditions for the maximum number of newly designed modules, LLM calls, wall-clock time and rounds without improvement. Given multitask data splits, *LLMSearchAgent* runs the guided decoder search, evaluates candidate decoders on the specified validation sets and writes the selected architectures and scores back to the history file. This provides a reproducible entry point for rerunning LLM–driven AAS with fixed data splits, module descriptions and search history.

```
    from bcijelly import LLMSearchAgent
    llm_agent = LLMSearchAgent(
                max_design_attempts = 3,
                max_llm_calls = 25,
                max_wall_clock_min = 60,
                no_improve_rounds = 2)
    multitask_data = llm_agent.split_data(
```

```
                    data_path = "./data/invasive",
                    split_mode = "csv")
        llm_agent.search(
                    multitask_data,
                    task_file = "./docs/task.md",
                    module_files = "./docs/modules/",
                    history_file = "./docs/history.json")
```

## Hardware platforms

**GPU, Lynxi and TaiBai neuromorphic chip.** The BCIJelly deployment pipeline targets three hardware platforms, namely conventional GPUs, the commercially available Lynxi HE200 neuromorphic chip and TaiBai, our recently developed neuromorphic chip. GPU experiments were performed on an NVIDIA RTX A6000 with CUDA 12.4 and PyTorch 2.11.0. Lynxi HE200 is a commercially available neuromorphic chip that supports both artificial and spiking neural network deployment, with compilation and deployment performed through the BCIJelly *toChip* pipeline, which wraps the vendor software stack. TaiBai is a fully programmable, many-core brain-inspired processor organized as an 11×12 array of cortical column cores communicating over a 2D mesh interconnect, supporting flexible spiking neuron models and on-chip learning. All the TaiBai results reported in this work were obtained on its FPGA prototype implementation, with compilation and deployment performed through the BCIJelly *toChip* pipeline. Native operator coverage of these platforms (**Fig. 5b**) was documented from official software stack and hardware specifications.

**Model compilation and deployment.** The BCIJelly deployment pipeline takes a trained PyTorch model checkpoint as input and produces a binary executable on the target chip in two stages, namely model conversion and compilation. In the conversion stage, PyTorch operators are translated into an intermediate representation tailored to the target chip. For ANN models, conversion is direct. For the SNN models, an additional step converts the PyTorch model into its spike-based equivalent using the SpikingJelly library[49] (**Fig. 5a**). In the compilation stage, the resulting graph is compiled into instructions specific to each chip through operator fusion and graph placement optimization. The pipeline is exposed through a single function called *bcijelly.toChip*, with the mode argument controlling the deployment path (**Fig. 1b**). Currently supported toolchains are Lynxi (ANN and SNN) and TaiBai (SNN only).

For the SNN models, sequences of the Conv2D, BatchNorm2D and LIF layers in the computational graph are extracted and fused into a single spike-Conv2D operator that maps directly onto the chip's native spike convolution instruction (**Fig. 5d**). The fused graph is then mapped onto the core array of the chip using the compilation framework of TaiBai, which minimizes the number of cores used subject to per-core memory and fan-in/fan-out constraints. The compiled models were verified by comparing outputs against the PyTorch reference, and deployment preserved decoding accuracy across the platforms (**Fig. 5c**).

**Power and resource measurements.** Power consumption for the MFSNN model was measured during decoder inference on each platform. The GPU power was recorded through the pynvml API. Lynxi HE200 power was obtained through the vendor-provided monitoring tool (*lynxi-smi*). The TaiBai power was estimated from its behavioral chip simulator. The power for each platform was averaged over approximately 200 held-out trials (**Fig. 5c**). The on-chip core count was measured as the number of cores allocated by the BCIJelly compilation pipeline to host the MFSNN model. The full BCIJelly optimization pipeline reduced the core count from 113 under default operator placement to 33 (**Fig. 5d**).

**Visualization software**

The BCIJelly visualization platform is a standalone desktop application developed in C++ that uses the Qt 6 framework, and provides real-time visualization of neural recordings and decoding results during BCI experiments. Multichannel neural traces are rendered through OpenGL for efficient real-time display, with support for zoom and drag interactions. The recorded data are loaded in HDF5 format, and the raw acquisition files in rs6 or rhd format are converted to HDF5 by the platform.

Anatomical reference is provided through hierarchical 3D brain region models in OBJ mesh format, with species-specific atlases for human, macaque and mouse experiments. Channels can be selected either through the waveform panel or by clicking on the corresponding 3D brain region, with the two views linked through a shared channel-to-region mapping. Butterworth filtering is available for noise removal, with cutoff frequencies and filter order adjustable through the configuration panel, and multiple filters can be combined for sequential processing.

Decoders are loaded through the configuration panel by specifying the algorithm name and a TCP port. Each decoder runs as a separate backend process that receives windowed neural input from the platform and returns either a discrete reach direction (displayed on an eight-direction radial layout) or a continuous motion trajectory, with results overlaid on the displayed waveforms in real time. The current release ships with backend implementations for the MLP, GRU, LSTM, Transformer, LFADS, Stabilization and Cycle-GAN, and additional decoders can be integrated by registering new TCP backends through the configuration panel. The BCIJelly visualization platform is released as open source and is freely available at https://github.com/LiyuanHan/BCIJelly, together with a demonstration of its use (**Supplementary Video 1**).

**Extended Data Fig. 1:** Decoder performance across five regression datasets.

**Extended Data Fig. 2:** Architectures of the best decoders identified by AAS and their core-building algorithmic modules.

**Extended Data Fig. 3:** Architecture of the hardware-deployment cross-day regression decoder searchedd by AAS for the Jango dataset.

**Extended Data Fig. 4:** Deployment of three additional decoders on the Lynxi neuromorphic chip.

**Extended Data Fig. 5:** Architecture of the shared encoder.

**Extended Data Fig. 6:** Single-task decoding performance of closed-loop AAS with LLMs across five invasive BCI datasets.

**Extended Data Fig. 7:** Architecture of UniBCI.

**Supplementary Table 1:** Fixed steps for classification analyses and bin sizes for regression analyses across datasets.

**Supplementary Table 2:** Training and test data spans across datasets for single-day and cross-day classification analyses.

**Supplementary Table 3:** Performance of single-day classification across datasets.

**Supplementary Table 4:** Performance of cross-day classification across datasets.

**Supplementary Table 5:** Training and test data spans across datasets for single-day and cross-day regression analyses.

**Supplementary Table 6:** Performance of single-day regression across datasets (Part I).

**Supplementary Table 7:** Performance of single-day regression across datasets (Part II).

**Supplementary Table 8:** Performance of cross-day regression across datasets (Part I).

**Supplementary Table 9:** Performance of cross-day regression across datasets (Part II).

**Supplementary Table 10:** Parameter table of the AAS.

**Supplementary Table 11.** Training, validation and testing data spans across 5 datasets for single-task, multitask and cross-species decoding settings.

**Supplementary Table 12:** Pseudocode of the AAS.

**Supplementary Table 13:** Parameters of benchmark decoders (MLP, LSTM and Transformer) for classification and regression BCI tasks.

**Supplementary Table 14:** Parameters of benchmark decoders (NDT 1, NDT 2 and POYO) for classification and regression BCI tasks.

**Supplementary Table 15:** Parameters of benchmark decoders (LFADS and Stabilization) for classification and regression BCI tasks.

**Supplementary Table 16:** Parameters of benchmark decoders (Cycle-GAN and NoMAD) for classification and regression BCI tasks.

**Supplementary Table 17:** Parameters of benchmark decoders (Speech LSTM and Speech Transformer) for classification and regression BCI tasks.

**Supplementary Table 18:** Parameters of benchmark decoders (MSCFormer and MFSNN) for classification BCI tasks.

**Supplementary Table 19:** Parameters of benchmark decoders (KF, DFINE, FENet and seqVAE) for regression BCI tasks.

**Supplementary Table 20:** Shared encoder architecture in the LLM–driven AAS.

**Supplementary Note 1:** The parameters override those of some selected benchmark decoders across datasets.

**Supplementary Video 1:** Interactive visualization throughout the BCIJelly workflow.

**Supplementary Video 2**: Closed-loop AAS with the LLM (Claude Opus 4.7) for 5 BCI tasks.

**Supplementary File 1:** Paired statistical comparisons across pretrained models and AAS with LLMs.

**Supplementary File 2**: Per-day testing scores of the pretrained models and AAS with LLMs across all tasks.

**Supplementary File 3**: Single-day and cross-day performance of BCIJelly benchmark decoders on classification tasks.

**Supplementary File 4**: Single-day and cross-day performance of BCIJelly benchmark decoders on regression tasks.

**Acknowledgments** We thank Yu Song, Erjun Xiao, Boshi Zhao, and Qianpeng Li for helpful discussions and Haining Zhang and Chenxi Xu for their contributions to the website design. **Funding** This work was supported by the Brain Science and Brain-like Intelligence Technology - National Science and Technology Major Project (2025ZD0217200), Strategic Priority Research Program of the Chinese Academy of Sciences (Grant No. XDB1010302), CAS Project for Young Scientists in Basic

Research (YSBR-116), Youth Innovation Promotion Association CAS, Shanghai Leading Talent Program of Eastern Talent Plan, the Lingang Laboratory Fund (Grant No. LG-GG-202402-06-07, LGL-1987-09), the Shanghai Municipal Science and Technology Project (Grant No. 25ZR1401370, 25LN3200400), and the Special Support Project of Guangdong Province (Grant No.0720240209). The numerical calculations in this study were carried out on the ORISE Supercomputer. **Author contributions** T.Z. and B.X. designed the study. T.Z., L.H., X.Y, T.Y.Z., B.X., C.L. and M.P. performed the discussion and wrote the paper together. L.H., T.Z, B.H., and R.X. performed the experiments on the GPU. L.H., T.Y.Z., and Q.Y. constructed the BCIJelly python library. Y.Q. and L.C. performed the experimental tests on the neuromorphic chip. T.Z., L.H., and X.Y. concatenated the experimental figures. L.H., X.Y., T.Y.Z., and Q.Y collected public invasive BCI datasets and performed data preprocessing. Y.G., and M.P., designed the biological BCI paradigm and collected the neural signals recorded by the electrodes. **Competing interests** The authors declare that they have no competing interests.

**Data availability** The invasive BCI datasets used in this study are publicly available at https://huggingface.co/datasets/LiyuanHan/bcijelly-invasive.

**Code and software availability** The BCIJelly Python package is available from PyPI (https://pypi.org/project/bcijelly/) and can be installed via *pip install bcijelly*. Documentation and usage tutorials for BCIJelly are available at https://bcihub.github.io/BCIJelly-site. The visualization software is available via Google Drive at https://drive.google.com/drive/folders/1QP-nBIPxlg4mrJ7myQ6EIR U63Imnh2Ti?usp=sharing.

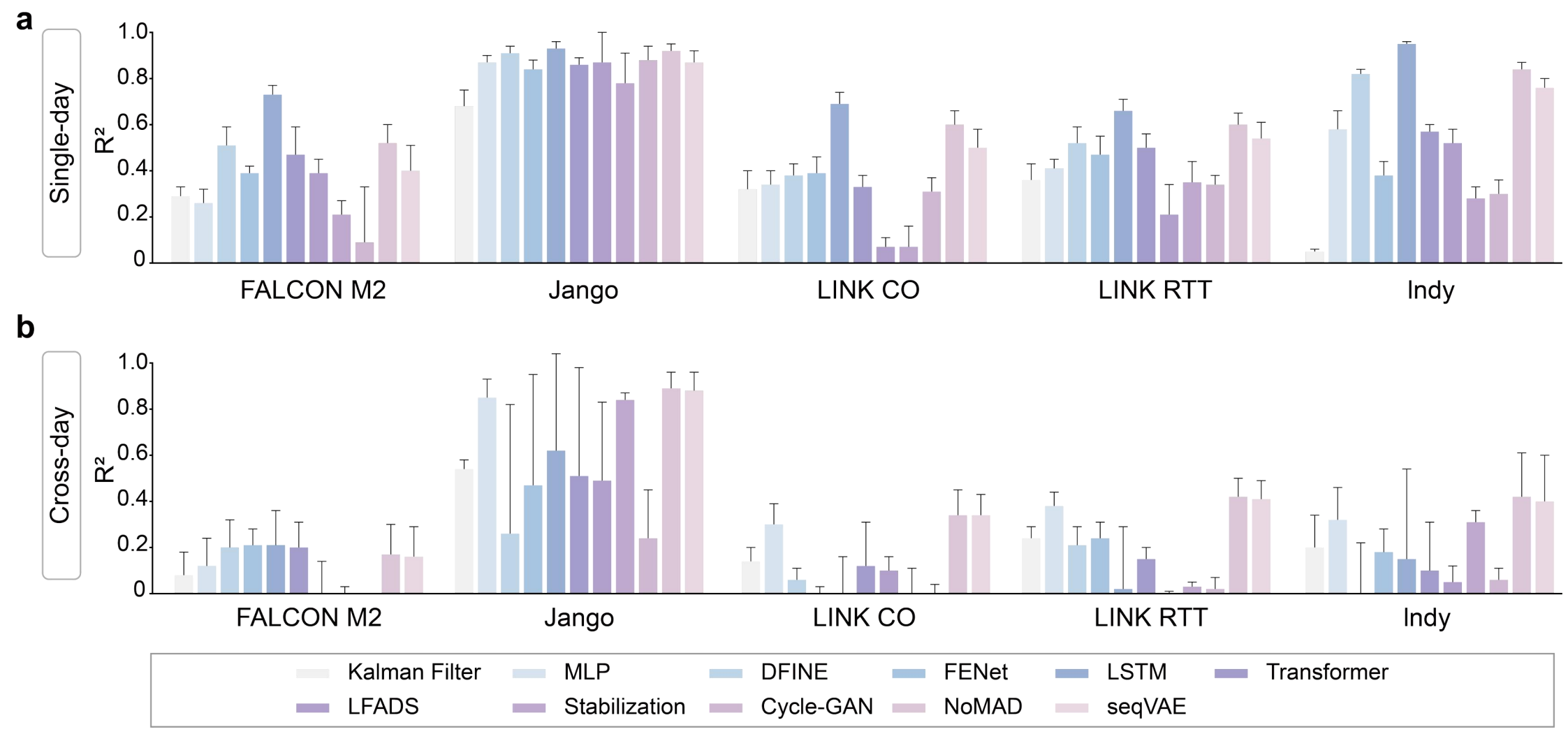

**Extended Data Fig. 1 | Decoder performance across five regression datasets.** Performance of 11 representative decoders (KF, MLP, DFINE, FENet, LSTM, Transformer, LFADS, Stabilization, Cycle-GAN, NoMAD and seqVAE) implemented in BCIJelly and benchmarked on five regression datasets recorded from macaques, comprising two center-out reaching tasks (Jango and LINK CO) and three random target tracking tasks (FALCON M2, LINK RTT and Indy), in which continuous behavioral trajectories are decoded point by point from neural activity. Top, single-day decoding, where each decoder is trained and tested on the same recording day. Bottom, cross-day decoding, where each decoder is trained on a subset of recording days and tested on days held out from training. The performance is reported as the coefficient of determination ($R^2$), with bars indicating the mean across test days and error bars indicating one standard deviation. No single decoder is dominant across datasets or across the two evaluation settings, the best-performing decoder differs from one dataset to the next, and the cross-day performance is consistently lower than the single-day performance, reflecting the challenge of generalizing decoders across recording days.

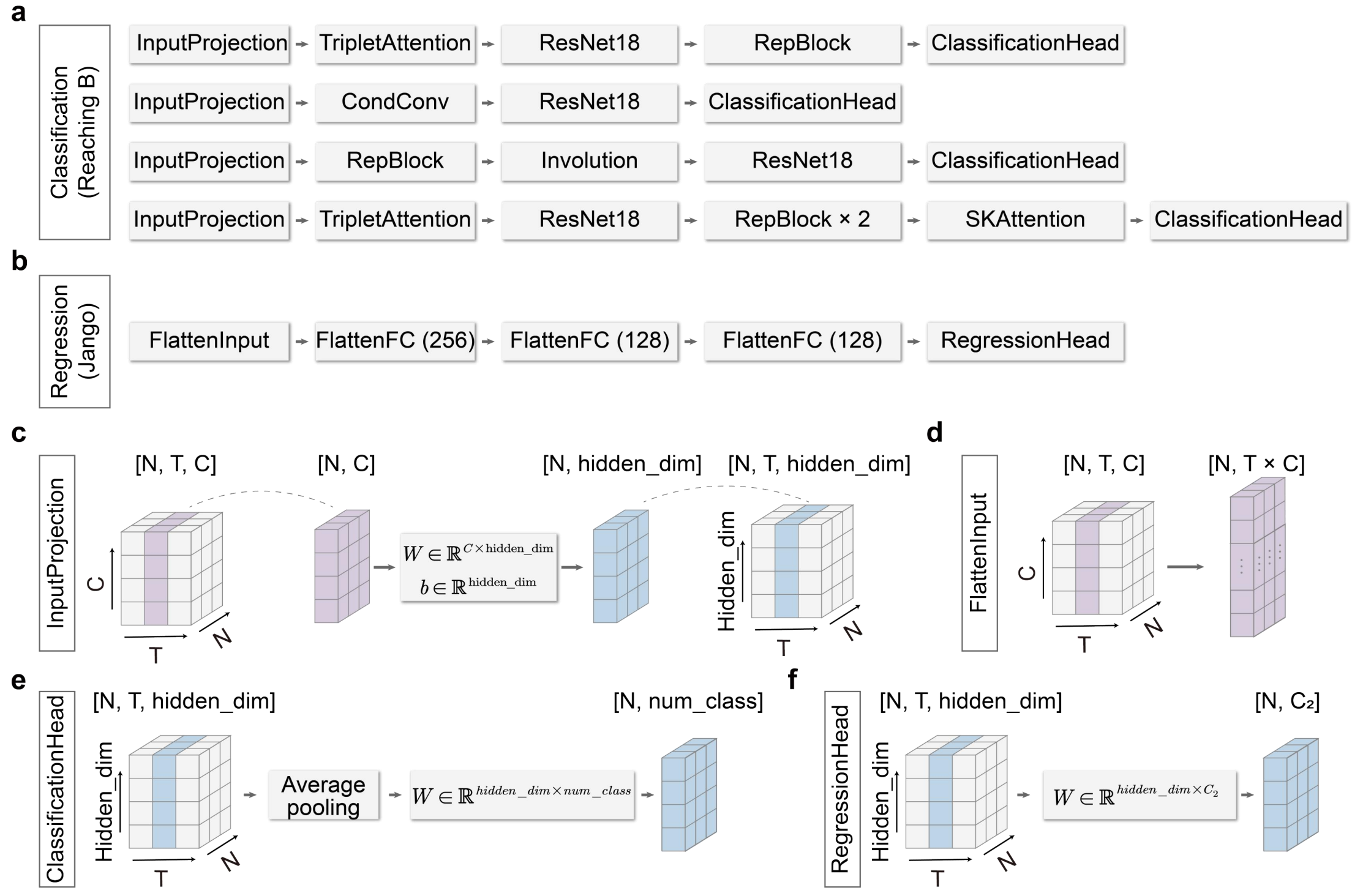

**Extended Data Fig. 2 | Architectures of the best decoders identified by AAS and their core-building algorithmic modules. a,** Model architecture corresponding to the best architecture search result for the classification task in Fig. 3d, which is an ensemble of four subnetworks. Each subnetwork shares the same overall structure, starting from an InputProjection module and ending in a ClassificationHead, but differs in its intermediate feature-extraction modules, which combine components such as TripletAttention, CondConv, RepBlock, Involution, SKAttention and a ResNet18 backbone. The four subnetworks are combined into the final classification decoder. **b,** Model architecture corresponding to the best architecture search result for the regression task in Fig. 3d, consisting of a FlattenInput module followed by three fully connected layers (FlattenFC with 256, 128 and 128 units) and a RegressionHead. **c,** The InputProjection module, which maps the input neural tensor of shape (C, T, N), where C is the number of channels, T is the number of time steps and N is the number of trials, into a hidden representation through a learned linear transformation with weight W and bias b. **d,** The FlattenInput module, which reshapes the neural input for the fully connected regression network. e, The ClassificationHead, which applies average pooling over the temporal dimension followed by a linear projection (weight

W of shape *hidden_dim* × *num_class*) to produce class scores. **f,** The RegressionHead, which applies a linear projection (weight W of shape *hidden_dim* × $C_2$, where $C_2$ is the dimensionality of the continuous output) to produce the regression output.

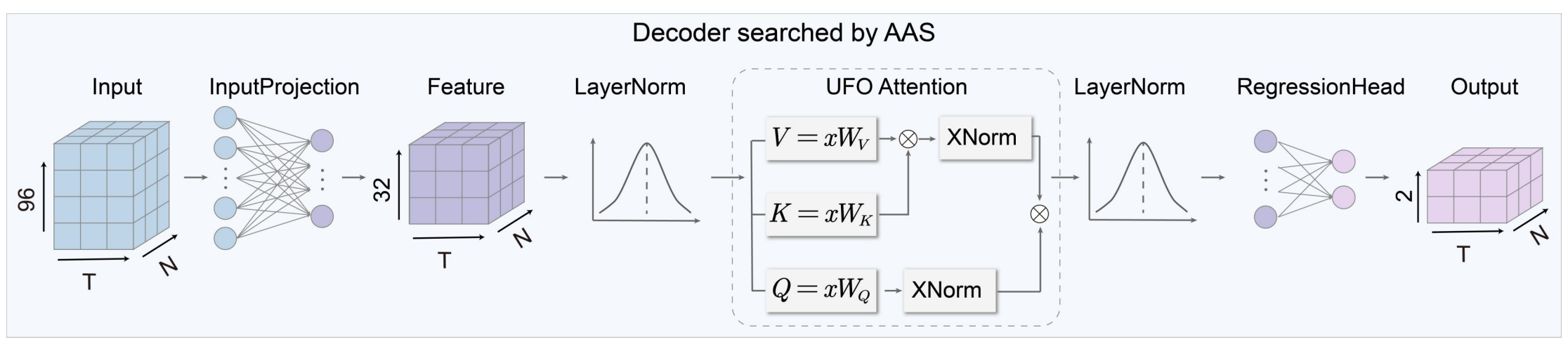

**Extended Data Fig. 3 | Architecture of the hardware-deployment cross-day regression decoder searched by AAS for the Jango dataset.** The decoder takes a neural input tensor of shape [N, T, 96], where N is the number of trials, T is the number of time steps and 96 is the number of channels, and produces a continuous output of shape [N, T, 2], corresponding to the two-dimensional (x, y) position. The input is first mapped into a hidden representation by a fully connected projection and passed through a nonlinear activation. The core of the network is a UFO attention module identified by the architecture search (dashed box), in which the hidden representation x is projected into the query, key and value through learned weights. The key and value branches are each followed by a cross-normalization step (XNorm), and the branches are combined through successive matrix multiplication to produce the attended representation. A further nonlinear activation and a fully connected projection then map this representation to the two-dimensional output. This decoder generalizes across recording days, achieving successful cross-day decoding on the Jango dataset.

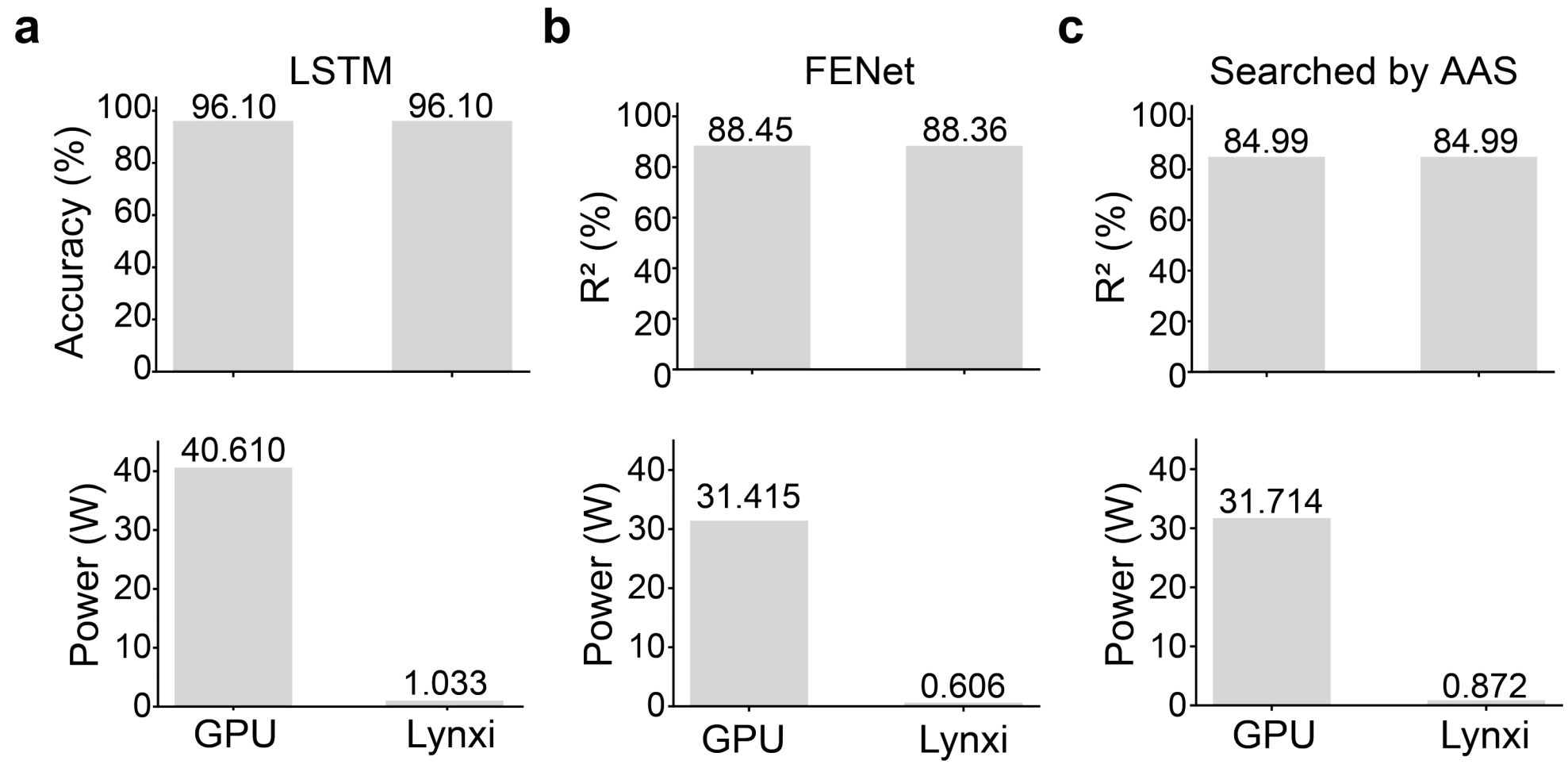

**Extended Data Fig. 4 | Deployment of three additional decoders on the Lynxi neuromorphic chip.** Decoding performance and power consumption of three decoders deployed on a GPU and on the Lynxi neuromorphic chip, following the same deployment procedure (**Fig. 5a**). **a**, LSTM, a decoder collected in BCIJelly, evaluated on a classification task using the Finger PPC dataset. The decoding accuracy is 96.10% on both the GPU and Lynxi (top), while power consumption is reduced from 40.610 W to 1.033 W (approximately 39-fold, bottom). **b**, FENet, also collected in BCIJelly, evaluated on a regression task using the Jango dataset. The decoding performance ($R^2$) is 0.8845 on the GPU and 0.8836 on Lynxi (top), while power consumption is reduced from 31.415 W to 0.606 W (approximately 52-fold, bottom). **c**, A decoder searched by AAS, was evaluated on the same regression task. $R^2$ is 0.8499 on both platforms (top), while power consumption is reduced from 31.714 W to 0.872 W (roughly 36-fold, bottom). Across all three decoders, the decoding performance on the GPU and Lynxi is essentially identical, confirming that operator conversion preserves model performance during deployment. These results demonstrate that the BCIJelly *toChip* pipeline generalizes across both collected and automatically generated decoders.

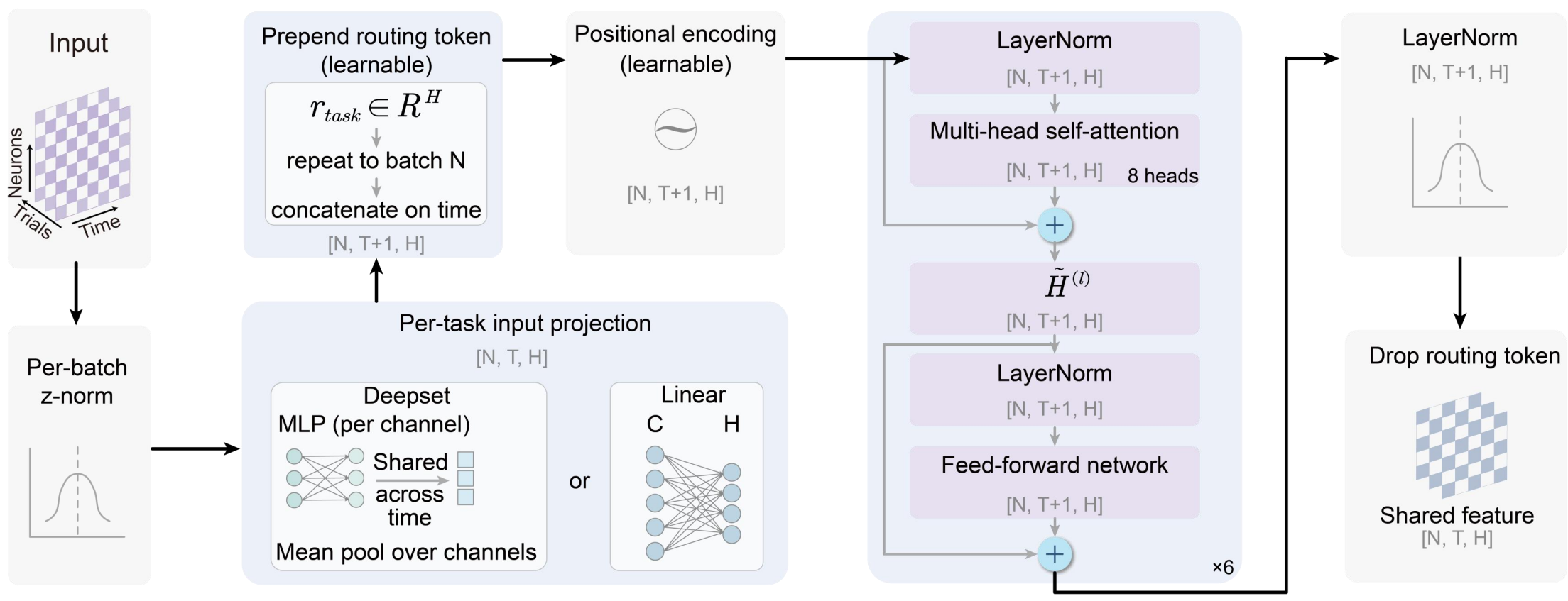

**Extended Data Fig. 5 | Architecture of the shared encoder.** The shared encoder maps a neural input tensor of shape [N, T, C], where N is the number of trials, T is the number of time steps and C is the number of channels, to a shared feature representation of shape [N, T, H], where H is the hidden dimensionality. The input is first normalized per batch (per-batch z-norm) and passed through a per-task input projection, implemented either as a permutation-invariant Deepset module (a shared per-channel MLP followed by mean pooling over channels) or as a linear layer, producing a representation of shape [N, T, H]. A learnable global token $r_{task} \in R^H$ is prepended along the time dimension, yielding a sequence of shapes [N, T+1, H], to which a learnable positional encoding is added. This token allows all the time steps to exchange information through self-attention and is discarded after the final block. The sequence is processed by a stack of six Transformer blocks, each consisting of layer normalization, multi-head self-attention (8 heads) and a feed-forward network with residual connections, where $H^{(l)}$ denotes the hidden state after the $l^{th}$ block. After a final layer normalization, the global token is dropped to yield the per-time-step shared feature of shape [N, T, H], which is then passed to the task-specific decoders.

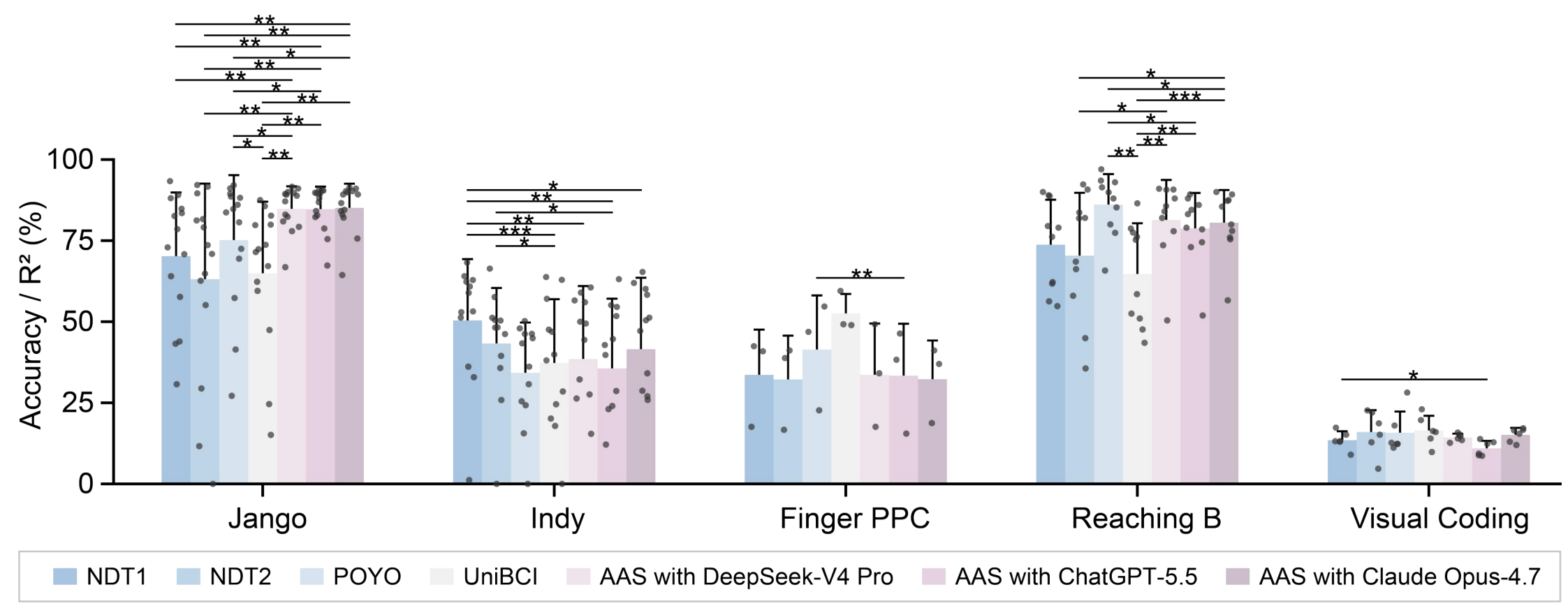

**Extended Data Fig. 6 | Single-task decoding performance of closed-loop AAS with LLMs across five invasive BCI datasets.** Each decoder is trained and evaluated separately on each of the five datasets (Jango, Indy, Finger PPC, Reaching B and Visual Coding) without joint training across datasets or species, in contrast to the cross-species joint-training settings in Fig. 4b. Models and the bar, dot and error bar conventions are as in Fig. 4b. Accuracy is used for the classification datasets (Finger PPC, Reaching B and Visual Coding), and the coefficient of determination ($R^2$, shown as a percentage) is used for the regression datasets (Jango and Indy). Even in this single-task setting, the decoders searched by LLM–driven AAS remained competitive with the state-of-the-art pretrained models, exceeding them on Jango and performing comparably on Finger PPC, Reaching B and Visual Coding, whereas NDT1 retained an advantage on Indy. UniBCI was competitive on Finger PPC and Visual Coding, where it reached the highest mean performance. Statistical significance was assessed using a paired *t* test. $^*P < 0.05$, $^{**}P < 0.01$, $^{***}P < 0.001$.

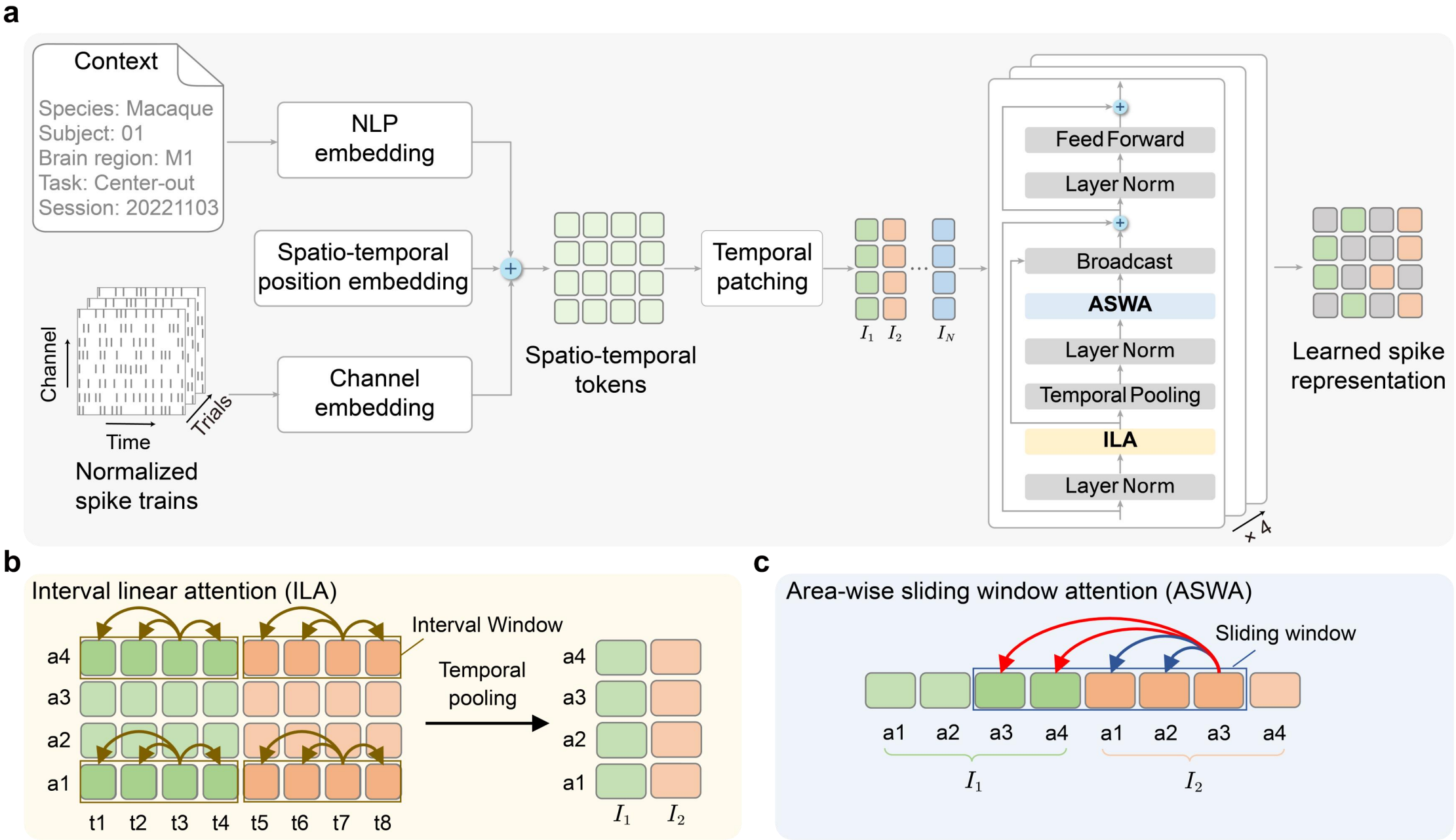

**Extended Data Fig. 7 | Architecture of UniBCI. a**, UniBCI combines context-aware tokenization with a stack of interval-area attention layers. Normalized spike trains are tokenized together with contextual information (species, subject, brain region, task and session), with an NLP embedding, a channel embedding and a spatio-temporal position embedding summed to form spatio-temporal tokens, which are split by temporal patching into successive intervals ($I_1$ to $I_N$). The tokens are then processed by L stacked layers, each applying interval linear attention (ILA), temporal pooling, area-wise sliding window attention (ASWA) and a feed-forward network, with the pooled interval representation broadcast back across time steps before being added to the sequence, producing a learned spike representation. **b**, Within each interval window, the ILA applies a linear-complexity attention over time to capture temporal features within a short interval, followed by temporal pooling. **c**, A sliding window spanning neighboring intervals captures longer-range temporal dependencies across intervals.

# Supplementary Tables

## Supplementary Table 1. Fixed steps for classification analyses and bin sizes for regression analyses across datasets.

| Datasets | Fixed steps for classification analyses | Bin size for regression analyses (ms) |
|---|---|---|
| FALCON M1-A | 100 | - |
| Finger RL | 50 | - |
| Finger PPC | 50 | - |
| Jango | 30 | 50 |
| Reaching C-J-M-T | 50 | - |
| Reaching C-M | 30 | - |
| Reaching B | 50 | - |
| Grasp P-G | 30 | - |
| Maze N-J | - | 30 |
| Visual Coding | 50 | - |
| Visual Grating | 300 | - |
| Music | 50 | - |
| Emotion | 30 | - |
| Speech | 906 | - |
| FALCON M2 | - | 20 |
| LINK CO | - | 20 |
| LINK RTT | - | 20 |
| Indy | - | 20 |

## Supplementary Table 2. Training and test data spans across datasets for single-day and cross-day classification analyses.

| Datasets | Single-day decoding | | Cross-day decoding | | | |
|---|---|---|---|---|---|---|
| | Training / test data span | Training / test days | Training data span | Training days | Test data span | Test days |
| Finger PPC | 20180910 ~ 20181022 | 10 | 20180910 ~ 20181012 | 7 | 20181017 ~ 20181022 | 2 |
| Reaching C-J-M-T | 20131003 ~ 20161021 | 53 | 20150629 ~ 20150707 | 6 | 20150709 ~ 20150710 | 2 |
| Reaching B | Day 1 ~ Day 37 | 37 | Day 1 ~ Day 26 | 26 | Day 30 ~ Day 37 | 8 |
| Jango | 20150730 ~ 20151102 | 20 | 20150730 ~ 20150828 | 14 | 20150906 ~ 20151102 | 4 |
| Visual Grating | 0805 ~ 1015 | 10 | 0722 ~ 0804 | 5 | 0806 | 1 |
| Visual Coding | 715093703 ~ 799864342 | 32 | 715093703 ~ 760345702 | 22 | 762602078 ~ 799864342 | 7 |
| Music | Patient 00 ~ Patient 28 | 29 | Patient 00 ~ Patient 20 | 21 | Patient 23 ~ Patient 28 | 6 |
| Emotion | 20100708 ~ 20110706 | 9 | 20100708 ~ 20110517 | 6 | 20110705 ~ 20110706 | 2 |
| Speech | Day 0 ~ Day 23 | 24 | Day 0 ~ Day18 | 19 | Day19 ~ Day23 | 5 |

## Supplementary Table 3. Performance of single-day classification across datasets.

| Datasets | Performance of single-day decoding | | | | | | | | |
|---|---|---|---|---|---|---|---|---|---|
| | MLP | LSTM | Transformer | MSCFormer | LFADS | Stabilization | Cycle-GAN | NoMAD | MFSNN |
| Finger PPC | 0.90 ± 0.04 | 0.90 ± 0.09 | 0.88 ± 0.11 | 0.77 ± 0.06 | 0.88 ± 0.06 | 0.69 ± 0.09 | 0.84 ± 0.12 | 0.84 ± 0.06 | 0.72 ± 0.10 |
| Reaching C-J-M-T | 0.78 ± 0.11 | 0.67 ± 0.16 | 0.89 ± 0.11 | 0.84 ± 0.09 | 0.72 ± 0.08 | 0.58 ± 0.08 | 0.70 ± 0.13 | 0.72 ± 0.23 | 0.33 ± 0.19 |
| Reaching B | 0.85 ± 0.10 | 0.83 ± 0.13 | 0.90 ± 0.12 | 0.88 ± 0.11 | 0.80 ± 0.13 | 0.81 ± 0.12 | 0.82 ± 0.12 | 0.87 ± 0.12 | 0.68 ± 0.12 |
| Jango | 0.95± 0.04 | 0.93 ± 0.07 | 0.95 ± 0.09 | 0.86 ± 0.06 | 0.96 ± 0.04 | 0.74 ± 0.15 | 0.93 ± 0.10 | 0.84 ± 0.11 | 0.77 ± 0.09 |
| Visual Grating | 0.77 ± 0.07 | 0.68 ± 0.11 | 0.63 ± 0.11 | 0.67 ± 0.11 | 0.58 ± 0.08 | 0.76 ± 0.06 | 0.78 ± 0.09 | 0.68 ± 0.11 | 0.50 ± 0.01 |
| Visual Coding | 0.73 ± 0.13 | 0.78 ± 0.13 | 0.80 ± 0.13 | 0.44 ± 0.12 | 0.67 ± 0.14 | 0.49 ± 0.24 | 0.44 ± 0.30 | 0.73 ± 0.14 | 0.51 ± 0.15 |
| Music | 0.86 ± 0.07 | 0.88 ± 0.06 | 0.87 ± 0.06 | 0.83 ± 0.00 | 0.84 ± 0.06 | 0.80 ± 0.10 | 0.83 ± 0.06 | 0.83 ± 0.08 | 0.85 ± 0.04 |
| Emotion | 0.38 ± 0.06 | 0.34 ± 0.05 | 0.34 ± 0.05 | 0.37 ± 0.06 | 0.36 ± 0.06 | 0.38 ± 0.05 | 0.37 ± 0.04 | 0.37 ± 0.06 | 0.33 ± 0.04 |
| Speech | - | 0.48 ± 0.15 | 0.29 ± 0.19 | - | - | - | - | - | - |

## Supplementary Table 4. Performance of cross-day classification across datasets.

| Datasets | Performance of cross-day decoding | | | | | | | | |
|---|---|---|---|---|---|---|---|---|---|
| | MLP | LSTM | Transformer | MSCFormer | LFADS | Stabilization | Cycle-GAN | NoMAD | MFSNN |
| Finger PPC | 0.39 ± 0.07 | 0.32 ± 0.06 | 0.29 ± 0.07 | 0.41 ± 0.05 | 0.32 ± 0.05 | 0.34 ± 0.07 | 0.28 ± 0.12 | 0.51 ± 0.04 | 0.24 ± 0.09 |
| Reaching C-J-M-T | 0.13 ± 0.05 | 0.12 ± 0.02 | 0.13 ± 0.02 | 0.17 ± 0.05 | 0.21 ± 0.05 | 0.29 ± 0.05 | 0.13 ± 0.02 | 0.17 ± 0.05 | 0.16 ± 0.05 |
| Reaching B | 0.67 ± 0.14 | 0.55 ± 0.14 | 0.42 ± 0.14 | 0.77 ± 0.07 | 0.53 ± 0.12 | 0.77 ± 0.07 | 0.68 ± 0.13 | 0.80 ± 0.09 | 0.25 ± 0.10 |
| Jango | 0.48 ± 0.23 | 0.41 ± 0.07 | 0.51 ± 0.09 | 0.90 ± 0.05 | 0.51 ± 0.02 | 0.73 ± 0.06 | 0.41 ± 0.19 | 0.90 ± 0.01 | 0.63 ± 0.18 |
| Visual Grating | 0.77 ± 0.00 | 0.64 ± 0.02 | 0.68 ± 0.00 | 0.74 ± 0.02 | 0.59 ± 0.02 | 0.71 ± 0.02 | 0.54 ± 0.06 | 0.70 ± 0.01 | 0.50 ± 0.00 |
| Visual Coding | 0.12 ± 0.05 | 0.13 ± 0.01 | 0.12 ± 0.02 | 0.12 ± 0.03 | 0.14 ± 0.01 | 0.11 ± 0.01 | 0.15 ± 0.01 | 0.13 ± 0.05 | 0.13 ± 0.03 |
| Music | 0.82 ± 0.00 | 0.82 ± 0.04 | 0.81 ± 0.04 | 0.82 ± 0.00 | 0.82 ± 0.00 | 0.75 ± 0.08 | 0.80 ± 0.01 | 0.81 ± 0.04 | 0.82 ± 0.00 |
| Emotion | 0.36 ± 0.02 | 0.35 ± 0.02 | 0.33 ± 0.02 | 0.35 ± 0.02 | 0.35 ± 0.04 | 0.38 ± 0.01 | 0.35 ± 0.02 | 0.34 ± 0.02 | 0.33 ± 0.00 |
| Speech | - | 0.54 ± 0.16 | 0.48 ± 0.17 | - | - | - | - | - | - |

## Supplementary Table 5. Training and test data spans across datasets for single-day and cross-day regression analyses.

| Datasets | Single-day decoding | | Cross-day decoding | | | |
|---|---|---|---|---|---|---|
| | Training/test data span | Training/test days | Training data span | Training days | Test data span | Test days |
| FALCON M2 | 20201019 ~ 20201027 | 6 | 20201019 ~ 20201020 | 4 | 20201030 ~ 20201124 | 6 |

| Jango | 20150730 ~ 20151102 | 10 | 20150730 ~ 20150828 | 14 | 20150906 ~ 20151102 | 4 |
|---|---|---|---|---|---|---|
| LINK CO | 20210104 ~ 20210601 | 25 | 20210104 ~ 20210111 | 5 | 20210225 ~ 20210601 | 18 |
| LINK RTT | 20210305 ~ 20210703 | 25 | 20210305 ~ 20210322 | 5 | 20210412 ~ 20210703 | 18 |
| Indy | 20160418 ~ 20161026 | 25 | 20160418 ~ 20160622 | 5 | 20160630 ~ 20161026 | 18 |

**Supplementary Table 6. Performance of single-day regression across datasets (Part I).**

| **Datasets** | Performance of single-day decoding | | | | | |
|---|---|---|---|---|---|---|
| | KF | MLP | LSTM | Transformer | DFINE | FENet |
| FALCON M2 | 0.29 ± 0.04 | 0.26 ± 0.06 | 0.73 ± 0.04 | 0.47 ± 0.12 | 0.51 ± 0.08 | 0.39 ± 0.03 |
| Jango | 0.68 ± 0.07 | 0.87 ± 0.03 | 0.93 ± 0.03 | 0.86 ± 0.03 | 0.91 ± 0.03 | 0.84 ± 0.04 |
| LINK CO | 0.32 ± 0.08 | 0.34 ± 0.06 | 0.69 ± 0.05 | 0.33 ± 0.05 | 0.38 ± 0.05 | 0.39 ± 0.07 |
| LINK RTT | 0.36 ± 0.07 | 0.41 ± 0.04 | 0.66 ± 0.05 | 0.50 ± 0.06 | 0.52 ± 0.07 | 0.47 ± 0.08 |
| Indy | 0.05 ± 0.01 | 0.58 ± 0.08 | 0.95 ± 0.01 | 0.57 ± 0.03 | 0.82 ± 0.02 | 0.38 ± 0.06 |

**Supplementary Table 7. Performance of single-day regression across datasets (Part II).**

| **Datasets** | Performance of single-day decoding | | | | |
|---|---|---|---|---|---|
| | LFADS | Stabilization | Cycle-GAN | NoMAD | seqVAE |
| FALCON M2 | 0.39 ± 0.06 | 0.21 ± 0.06 | 0.09 ± 0.24 | 0.52 ± 0.08 | 0.40 ± 0.11 |
| Jango | 0.87 ± 0.13 | 0.78 ± 0.13 | 0.88 ± 0.06 | 0.92 ± 0.03 | 0.87 ± 0.05 |
| LINK CO | 0.07 ± 0.04 | 0.07 ± 0.09 | 0.31 ± 0.06 | 0.60 ± 0.06 | 0.50 ± 0.08 |
| LINK RTT | 0.21 ± 0.13 | 0.35 ± 0.09 | 0.34 ± 0.04 | 0.60 ± 0.05 | 0.54 ± 0.07 |
| Indy | 0.52 ± 0.06 | 0.28 ± 0.05 | 0.30 ± 0.06 | 0.84 ± 0.03 | 0.76 ± 0.04 |

**Supplementary Table 8. Performance of cross-day regression across datasets (Part I).**

| **Datasets** | Performance of cross-day decoding | | | | | |
|---|---|---|---|---|---|---|
| | KF | MLP | DFINE | FENet | LSTM | Transformer |
| FALCON M2 | 0.08 ± 0.10 | 0.12 ± 0.12 | 0.20 ± 0.12 | 0.21 ± 0.07 | 0.21 ± 0.15 | 0.20 ± 0.11 |
| Jango | 0.54 ± 0.04 | 0.85 ± 0.08 | 0.26 ± 0.56 | 0.47 ± 0.48 | 0.62 ± 0.42 | 0.51 ± 0.47 |
| LINK CO | 0.14 ± 0.06 | 0.30 ± 0.09 | 0.06 ± 0.05 | 0.00 ± 0.03 | -0.02 ± 0.18 | 0.12 ± 0.19 |
| LINK RTT | 0.24 ± 0.05 | 0.38 ± 0.06 | 0.21 ± 0.08 | 0.24 ± 0.07 | 0.02 ± 0.27 | 0.15 ± 0.05 |

| Indy | 0.20 ± 0.14 | 0.32 ± 0.14 | -0.07 ± 0.29 | 0.18 ± 0.10 | 0.15 ± 0.39 | 0.10 ± 0.21 |
|---|---|---|---|---|---|---|

**Supplementary Table 9. Performance of cross-day regression across datasets (Part II).**

| Datasets | Performance of cross-day decoding | | | | |
|---|---|---|---|---|---|
| | LFADS | Stabilization | Cycle-GAN | NoMAD | seqVAE |
| FALCON M2 | -0.10 ± 0.24 | -0.05 ± 0.08 | -0.12 ± 0.10 | 0.17 ± 0.13 | 0.16 ± 0.13 |
| Jango | 0.49 ± 0.34 | 0.84 ± 0.03 | 0.24 ± 0.21 | 0.89 ± 0.07 | 0.88 ± 0.08 |
| LINK CO | 0.10 ± 0.06 | -0.06 ± 0.17 | -0.04 ± 0.08 | 0.34 ± 0.11 | 0.34 ± 0.09 |
| LINK RTT | -0.00 ± 0.01 | 0.03 ± 0.02 | 0.02 ± 0.05 | 0.42 ± 0.08 | 0.41 ± 0.08 |
| Indy | 0.05 ± 0.07 | 0.31 ± 0.05 | 0.06 ± 0.05 | 0.42 ± 0.19 | 0.40 ± 0.20 |

**Supplementary Table 10. Parameter table of the AAS.**

| Category | Parameter | Default | Classification demo (Reaching B) | Regression Demo (Jango) |
|---|---|---|---|---|
| Data and task | task_type | None | classification | regression |
| Model skeleton | hidden_dim | 64 | 96 | 48 |
| | max_depth | 4 | 4 | 3 |
| Search budget | max_models_per_round | 8 | 4 | 6 |
| | max_search_rounds | 4 | 3 | 3 |
| Candidate filtering | val_threshold | 0.75 | 0.78 | 0.78 |
| | retain_top_k | 6 | 4 | 4 |
| Ensemble selection | ensemble_top_m | 4 | 4 | 4 |
| | ensemble_method | auto | auto | mean |
| Sampling constraint | max_per_type | 2 | 2 | 2 |
| | allow_repeat | True | True | True |
| | min_distinct_types | 1 | 2 | 1 |
| | no_repeat_within_types | None | backbone | backbone |
| Training | batch_size | 16 | 16 | 16 |
| | epochs | 12 | 12 | 18 |
| | patience | 4 | 4 | 5 |
| | learning_rate | 1e-3 | 1e-3 | 8e-4 |
| | weight_decay | 1e-4 | 1e-4 | 1e-4 |
| | normalize_regression_target | True | True | True |

**Supplementary Table 11. Training, validation and testing data spans across 5 datasets for single-task, multitask and cross-species decoding settings.**

| Datasets | Species | Training data span | Validate data span | Test data span | Test days | Task type | Validate threshold |
|---|---|---|---|---|---|---|---|
| Finger PPC | Human | 20180910 ~ 20181012 | 20181012 | 20181015 ~ 20181022 | 3 | Classific ation | 0.5 |

| Visual Coding | Mouse | 715093703 ~ 762120172 | 762602078 | 763673393 ~ 799864342 | 6 sessions | Classification | 0.2 |
|---|---|---|---|---|---|---|---|
| Reaching B | macaque | 20221103 ~ 20221125 | 20221128 | 20221129 ~ 20221230 | 10 | Classification | 0.7 |
| Jango | macaque | 20150730 ~ 20150805 | 20150806 | 20150808 ~ 20151102 | 14 | Regression | 0.6 |
| Indy | macaque | 20160915 ~ 20161006 | 20161007 | 20161011 ~ 20161220 | 12 | Regression | 0.6 |

## Supplementary Table 12. Pseudocode of the AAS.

*SearchAgent* algorithm

**Input:**

Training data D_train = (X_train, y_train);

Validation data D_val = (X_val, y_val);

Search configuration:

max_models_per_round, max_search_rounds, max_depth,

val_threshold, retain_top_k, ensemble_top_m

**Output:**

Best combined model f_best;

Search summary containing retained candidates and the selected ensemble

1: Discover candidate modules from the local model library.

2: Initialize the retained candidate set P = empty.

3. For round r = 1 to max_search_rounds:

4: Sample up to max_models_per_round candidate architectures under the search constraints.

5: For each candidate architecture a:

6: Build a time-series model:

7: InputProjection -> searched module blocks -> task head.

8: Train the model on D_train.

9: Evaluate the model on D_val.

10: If the validation metric of a is at least val_threshold:

11: add a to P.

12: Keep only the top retain_top_k candidates in P according to validation metric.

13: If |P| >= retain_top_k:

14: stop the search.

15: If P is empty:

16: use the best successfully trained candidates as the fallback pool.

17: Select the top ensemble_top_m candidates from P.

18: Evaluate candidate ensembles on D_val:

19: classification: combine class probabilities by mean or stacking;

20: regression: average predicted trajectories.

21: Choose the ensemble with the best validation metric.

22: Return the selected ensemble as f_best.

23 // External evaluation:

24: Apply f_best to each held-out test day.

25: Report day-wise metric, average metric, median metric, and minimum metric.

**Supplementary Table 13. Parameters of benchmark decoders (MLP, LSTM and Transformer) for classification and regression BCI tasks.**

<table>
<tr><th colspan="2" rowspan="2">Parameters</th><th colspan="2">Classification task</th><th colspan="2">Regression task</th></tr>
<tr><th>Single-day</th><th>Cross-day</th><th>Single-day</th><th>Cross-day</th></tr>
<tr><td rowspan="12">MLP</td><td>hidden_dims</td><td colspan="4">[512, 256, 128]</td></tr>
<tr><td>input</td><td colspan="4">flatten(time x channels)</td></tr>
<tr><td>normalization</td><td colspan="4">training-set z-score</td></tr>
<tr><td>num_hidden_layers</td><td colspan="4">3</td></tr>
<tr><td>batch_size</td><td colspan="4">512</td></tr>
<tr><td>dropout</td><td>0.3</td><td>0.35</td><td>0.3</td><td>0.35</td></tr>
<tr><td>epochs</td><td colspan="4">80</td></tr>
<tr><td>lr</td><td>1e-3</td><td>8e-4</td><td>1e-3</td><td>8e-4</td></tr>
<tr><td>patience</td><td colspan="4">15</td></tr>
<tr><td>save_ckpt</td><td colspan="4">False</td></tr>
<tr><td>val_ratio</td><td colspan="4">0.2</td></tr>
<tr><td>weight_decay</td><td colspan="4">1e-4</td></tr>
<tr><td rowspan="16">LSTM</td><td>bidirectional</td><td colspan="4">True</td></tr>
<tr><td>hidden_size</td><td colspan="2">64</td><td colspan="2">128</td></tr>
<tr><td>num_layers</td><td colspan="4">2</td></tr>
<tr><td>pooling</td><td colspan="2">mean</td><td colspan="2">-</td></tr>
<tr><td>batch_size</td><td colspan="2">32</td><td colspan="2">128</td></tr>
<tr><td>calib_epochs</td><td colspan="2" rowspan="3">-</td><td rowspan="3">-</td><td>8</td></tr>
<tr><td>calib_lr</td><td>0.0001</td></tr>
<tr><td>calib_ratio</td><td>0.2</td></tr>
<tr><td>dropout</td><td>0.1</td><td>0.2</td><td>0.1</td><td>0.2</td></tr>
<tr><td>epochs</td><td>6</td><td>40</td><td>25</td><td>50</td></tr>
<tr><td>lr</td><td>0.001</td><td>0.0005</td><td>0.001</td><td>0.0005</td></tr>
<tr><td>max_calib_trials</td><td colspan="3" rowspan="2">-</td><td>30</td></tr>
<tr><td>min_calib_trials</td><td>5</td></tr>
<tr><td>patience</td><td>20</td><td>12</td><td rowspan="2">-</td><td>12</td></tr>
<tr><td>val_ratio</td><td>0.2</td><td>0.2</td><td>-</td></tr>
<tr><td>weight_decay</td><td>0</td><td>0.001</td><td>0</td><td>0.001</td></tr>
<tr><td rowspan="16">Transformer</td><td>d_model</td><td colspan="4">64</td></tr>
<tr><td>n_heads</td><td colspan="4">4</td></tr>
<tr><td>num_layers</td><td colspan="4">2</td></tr>
<tr><td>batch_size</td><td colspan="2">32</td><td>128</td><td>64</td></tr>
<tr><td>calib_epochs</td><td colspan="2" rowspan="3">-</td><td rowspan="3">-</td><td>8</td></tr>
<tr><td>calib_lr</td><td>0.0001</td></tr>
<tr><td>calib_ratio</td><td>0.2</td></tr>
<tr><td>dropout</td><td>0.1</td><td>0.3</td><td>0.1</td><td>0.3</td></tr>
<tr><td>epochs</td><td>100</td><td>60</td><td>40</td><td>80</td></tr>
<tr><td>lr</td><td colspan="2">0.0001</td><td>0.0001</td><td>7e-05</td></tr>
<tr><td>max_calib_trials</td><td colspan="2" rowspan="2">-</td><td rowspan="2">-</td><td>30</td></tr>
<tr><td>min_calib_trials</td><td>5</td></tr>
<tr><td>patience</td><td>20</td><td>15</td><td>-</td><td>15</td></tr>
<tr><td>val_ratio</td><td>0.2</td><td>0.2</td><td>-</td><td>-</td></tr>
<tr><td>weight_decay</td><td>0</td><td>0.001</td><td>0.0001</td><td>0.0005</td></tr>
</table>

**Supplementary Table 14. Parameters of benchmark decoders (NDT 1, NDT 2 and POYO) for classification and regression BCI tasks.**

| Parameters | | Classification task | | Regression task | |
|---|---|---|---|---|---|
| | | Single-day | Cross-day | Single-day | Cross-day |
| NDT 1 | d_model | 512 | | | |
| | n_heads | 8 | | | |
| | num_layers | 5 | | | |
| | batch_size | 32 | | 128 | 64 |
| | calib_epochs | - | | | 8 |
| | calib_lr | | | | 0.0001 |
| | calib_ratio | | | | 0.2 |
| | dropout | 0.4 | | | |
| | epochs | 100 | 60 | 40 | 80 |
| | lr | 0.0001 | | | |
| | max_calib_trials | - | | | 30 |
| | min_calib_trials | | | | 5 |
| | patience | 20 | 15 | 20 | 15 |
| | val_ratio | 0.2 | | - | |
| | weight_decay | 0 | 0.001 | 0.0001 | 0.0005 |
| NDT 2 | d_model | 128 | | | |
| | n_heads | 8 | | | |
| | num_layers | 5 | | | |
| | batch_size | 32 | | 128 | 64 |
| | calib_epochs | - | | | 8 |
| | calib_lr | | | | 0.0001 |
| | calib_ratio | | | | 0.2 |
| | dropout | 0.2 | | | |
| | epochs | 100 | 60 | 40 | 80 |
| | lr | 0.0001 | | | |
| | max_calib_trials | - | | | 30 |
| | min_calib_trials | | | | 5 |
| | patience | 20 | 15 | 20 | 15 |
| | val_ratio | 0.2 | | - | |
| | weight_decay | 0 | 0.001 | 0.0001 | 0.0005 |
| POYO | d_model | 64 | | | |
| | n_heads | 4 | | | |
| | num_layers | 2 | | | |
| | batch_size | 32 | | 128 | 64 |
| | calib_epochs | - | | | 8 |
| | calib_lr | | | | 0.0001 |
| | calib_ratio | | | | 0.2 |
| | dropout | 0.3 | | | |
| | epochs | 100 | 60 | 40 | 80 |
| | lr | 0.0001 | | | |
| | max_calib_trials | - | | | 30 |
| | min_calib_trials | | | | 5 |
| | patience | 20 | 15 | 20 | 15 |
| | val_ratio | 0.2 | | - | |
| | weight_decay | 0 | 0.001 | 0.0001 | 0.0005 |

**Supplementary Table 15. Parameters of benchmark decoders (LFADS and Stabilization) for classification and regression BCI tasks.**

| Parameters | Classification task | Regression task |
|---|---|---|

| | | Single-day | Cross-day | Single-day | Cross-day |
|---|---|---|---|---|---|
| LFADS | fac_dim | 40 | | | |
| | batch_size | 64 | | 256 | |
| | dropout | 0.02 | | | |
| | epochs | 50 | | 80 | 50 |
| | lr | 0.004 | | 0.003 | 0.004 |
| | patience | 10 | | 15 | 10 |
| | val_ratio | 0.2 | | | |
| | weight_decay | 0 | | | |
| | decoder_type | - | | ridge | |
| | seed | 42 | | | |
| Stabilization | baseline_ntrials | 100 | | | |
| | hidden_size | 512 | | - | |
| | hidden_sizes | - | | [128, 64] | |
| | n_latents | 36 | | 10 | |
| | pooling | flatten | | - | |
| | align_n | 150 | 150 | 60 | 60 |
| | align_th | 0.1 | 0.1 | 0.01 | 0.01 |
| | batch_size | 32 | 32 | 128 | 128 |
| | calib_ratio | 0.1 | 0.1 | 0.2 | 0.2 |
| | enable_alignment | False | True | - | - |
| | epochs | 300 | | | |
| | fa_ll_thresh | 1e-05 | | | |
| | fa_max_its | 100000 | | | |
| | fa_min_priv_var | 1 | | 0.1 | |
| | fa_n_restarts | 5 | | | |
| | fine_tune | - | | True | |
| | fine_tune_epochs | | | 100 | |
| | single_day_mode | | | - | full_adaptation |
| | lr | 0.001 | | | |
| | max_calib_trials | 30 | | - | |
| | min_calib_trials | 5 | | 1 | |
| | patience | 0 | | | |
| | val_ratio | 0.2 | | | |
| | weight_decay | 0.0001 | | 0 | |
| | decoder_type | - | | mlp | |
| | kalman_gain_max_steps | | | 100000 | |
| | kalman_gain_tol | | | 1e-08 | |
| | seed | 42 | | | |

## Supplementary Table 16. Parameters of benchmark decoders (Cycle-GAN and NoMAD) for classification and regression BCI tasks.

| **Parameters** | | **Classification task** | | **Regression task** | |
|---|---|---|---|---|---|
| | | Single-day | Cross-day | Single-day | Cross-day |
| Cycle-GAN | discriminator_hidden | 0 | | 256 | |
| | generator_hidden | 0 | | 256 | |
| | decoder_hidden | - | | [128] | - |
| | decoder_num_layers | | | 2 | |
| | batch_size | 32 | | 128 | |
| | calib_ratio | 0.1 | | 0.2 | |
| | cycle_weight | 5 | | 7.5 | |
| | d_lr | 0.01 | | | |
| | decoder_epochs | 100 | | 50 | 60 |
| | decoder_lr | 0.001 | | | |

<table>
<tr><td rowspan="16"></td><td>dropout_d</td><td colspan="4">0.2</td></tr>
<tr><td>dropout_decoder</td><td colspan="4">0.3</td></tr>
<tr><td>dropout_g</td><td colspan="4">0.2</td></tr>
<tr><td>enable_alignment</td><td>False</td><td>True</td><td colspan="2">True</td></tr>
<tr><td>epochs</td><td colspan="2">400</td><td>200</td><td>300</td></tr>
<tr><td>g_lr</td><td colspan="4">0.001</td></tr>
<tr><td>identity_weight</td><td colspan="2">5</td><td colspan="2">7.5</td></tr>
<tr><td>loss_type</td><td colspan="4">L1</td></tr>
<tr><td>max_calib_trials</td><td colspan="4">30</td></tr>
<tr><td>min_calib_trials</td><td colspan="4">5</td></tr>
<tr><td>patience</td><td colspan="4">10</td></tr>
<tr><td>source_ratio</td><td colspan="4">0.9</td></tr>
<tr><td>val_ratio</td><td colspan="4">0.2</td></tr>
<tr><td>calib_epochs</td><td colspan="2" rowspan="3">-</td><td>12</td><td rowspan="3">-</td></tr>
<tr><td>calib_lr</td><td>0.0001</td></tr>
<tr><td>decoder_type</td><td>LSTM</td></tr>
<tr><td rowspan="17">NoMAD</td><td>align_epochs</td><td></td><td>5</td><td></td><td>5</td></tr>
<tr><td>align_lr</td><td colspan="4">5e-4</td></tr>
<tr><td>align_weight</td><td colspan="4">0.1</td></tr>
<tr><td>batch_size</td><td>128</td><td>64</td><td>128</td><td>64</td></tr>
<tr><td>dropout</td><td>0.25</td><td>0.3</td><td>0.2</td><td>0.25</td></tr>
<tr><td>enc_dim</td><td colspan="4">64</td></tr>
<tr><td>epochs</td><td>60</td><td>70</td><td>40</td><td>50</td></tr>
<tr><td>gen_dim</td><td colspan="4">64</td></tr>
<tr><td>hidden_dim</td><td colspan="2">128</td><td colspan="2">96</td></tr>
<tr><td>lr</td><td colspan="2">8e-4</td><td colspan="2">7e-4</td></tr>
<tr><td>patience</td><td>10</td><td>12</td><td>8</td><td>10</td></tr>
<tr><td>pooling</td><td colspan="2">mean</td><td colspan="2">-</td></tr>
<tr><td>rec_weight</td><td colspan="4">1e-3</td></tr>
<tr><td>restore_best</td><td colspan="4">True</td></tr>
<tr><td>smooth_weight</td><td colspan="4">1e-3</td></tr>
<tr><td>val_ratio</td><td colspan="4">0.2</td></tr>
<tr><td>weight_decay</td><td colspan="4">1e-4</td></tr>
</table>

**Supplementary Table 17. Parameters of benchmark decoders (Speech LSTM and Speech Transformer) for classification and regression BCI tasks.**

<table>
<tr><th colspan="2" rowspan="2">Parameters</th><th colspan="2">Speech task</th></tr>
<tr><th>Single-day</th><th>Cross-day</th></tr>
<tr><td rowspan="16">Speech LSTM</td><td>bidirectional</td><td colspan="2">True</td></tr>
<tr><td>gaussian_smooth_width</td><td colspan="2">2</td></tr>
<tr><td>hidden_size</td><td colspan="2">512</td></tr>
<tr><td>kernel_len</td><td colspan="2">14</td></tr>
<tr><td>num_layers</td><td colspan="2">4</td></tr>
<tr><td>stride_len</td><td colspan="2">4</td></tr>
<tr><td>use_day_idx</td><td colspan="2">True</td></tr>
<tr><td>batch_size</td><td colspan="2">32</td></tr>
<tr><td>constant_offset_sd</td><td colspan="2">0.05</td></tr>
<tr><td>dropout</td><td colspan="2">0.3</td></tr>
<tr><td>epochs</td><td colspan="2">50</td></tr>
<tr><td>lr</td><td colspan="2">0.0005</td></tr>
<tr><td>patience</td><td colspan="2">10</td></tr>
<tr><td>val_ratio</td><td colspan="2">0.2</td></tr>
<tr><td>weight_decay</td><td colspan="2">0.0001</td></tr>
<tr><td>white_noise_sd</td><td colspan="2">0.2</td></tr>
<tr><td>Speech</td><td>d_model</td><td colspan="2">256</td></tr>
</table>

| Transformer | gaussian_smooth_width | 2 |
| --- | --- | --- |
| | kernel_len | 14 |
| | n_heads | 8 |
| | num_layers | 4 |
| | stride_len | 4 |
| | use_day_idx | False |
| | batch_size | 32 |
| | constant_offset_sd | 0.05 |
| | dropout | 0.2 |
| | epochs | 50 |
| | lr | 0.0003 |
| | patience | 10 |
| | val_ratio | 0.2 |
| | weight_decay | 0.0001 |
| | white_noise_sd | 0.2 |

**Supplementary Table 18. Parameters of benchmark decoders (MSCFormer and MFSNN) for classification BCI tasks.**

| Parameters | | Classification task | |
| --- | --- | --- | --- |
| | | Single-day | Cross-day |
| MSCFormer | depth | 5 | |
| | f1 | 16 | |
| | heads | 8 | |
| | max_tokens | 512 | |
| | normalize | z-score | |
| | target_tokens | 20 | |
| | batch_size | 32 | |
| | dropout | 0.5 | |
| | epochs | 80 | |
| | lr | 1e-3 | 8e-4 |
| | patience | 20 | 15 |
| | val_ratio | 0.2 | |
| | weight_decay | 0 | 1e-4 |
| MFSNN | attention | True | |
| | groups | 16 | |
| | pooling_kernel | 10 | |
| | steps | 20 | |
| | tau | 2.0 | |
| | use_spikingjelly | True | |
| | batch_size | 32 | |
| | dropout | 0.5 | |
| | epochs | 100 | |
| | eval_interval | 1 | |
| | loss_type | mse | |
| | lr | 1e-2 | |
| | patience | 20 | |
| | save_ckpt | False | |
| | val_ratio | 0.2 | |
| | weight_decay | 5e-4 | |

**Supplementary Table 19. Parameters of benchmark decoders (KF, DFINE, FENet and seqVAE) for regression BCI tasks.**

<table>
<tr><th colspan="2" rowspan="2">Parameters</th><th colspan="2">Regression task</th></tr>
<tr><th>Single-day</th><th>Cross-day</th></tr>
<tr><td rowspan="8">KF</td><td>observation_covariance</td><td colspan="2">diag</td></tr>
<tr><td>observation_noise_floor</td><td colspan="2">1e-5</td></tr>
<tr><td>process_noise_floor</td><td colspan="2">1e-6</td></tr>
<tr><td>ridge</td><td colspan="2">1e-3</td></tr>
<tr><td>val_ratio</td><td colspan="2">0.2</td></tr>
<tr><td>lag_steps</td><td>[-10, -5, -3, -1, 1, 3, 5, 10]</td><td>[-5, -1, 0, 1, 3, 5, 10]</td></tr>
<tr><td>include_velocity</td><td colspan="2">True</td></tr>
<tr><td>target_dim</td><td colspan="2">2</td></tr>
<tr><td rowspan="23">DFINE</td><td>dim_a</td><td>2</td><td>2</td></tr>
<tr><td>dim_x</td><td>2</td><td>2</td></tr>
<tr><td>hidden_layer_list</td><td>[32, 32, 32]</td><td>[20, 20, 20, 20]</td></tr>
<tr><td>hidden_layer_list_mapper</td><td>[32, 32]</td><td>[20, 20, 20]</td></tr>
<tr><td>activation</td><td colspan="2">tanh</td></tr>
<tr><td>activation_mapper</td><td colspan="2">tanh</td></tr>
<tr><td>batch_size</td><td colspan="2">128</td></tr>
<tr><td>calib_epochs</td><td colspan="2">10</td></tr>
<tr><td>calib_lr</td><td colspan="2">0.0001</td></tr>
<tr><td>calib_ratio</td><td colspan="2">0.2</td></tr>
<tr><td>calib_scope</td><td colspan="2">mapper</td></tr>
<tr><td>epochs</td><td colspan="2">70</td></tr>
<tr><td>grad_clip</td><td colspan="2">1</td></tr>
<tr><td>lr</td><td colspan="2">0.02</td></tr>
<tr><td>max_calib_trials</td><td colspan="2">30</td></tr>
<tr><td>min_calib_trials</td><td colspan="2">5</td></tr>
<tr><td>patience</td><td colspan="2">10</td></tr>
<tr><td>reduce</td><td colspan="2">sequence</td></tr>
<tr><td>restore_best</td><td colspan="2">True</td></tr>
<tr><td>scale_behv_recons</td><td>10</td><td>20</td></tr>
<tr><td>scale_l2</td><td colspan="2">0</td></tr>
<tr><td>steps_ahead</td><td>[1, 2, 3]</td><td>[1, 2, 3, 4]</td></tr>
<tr><td>val_ratio</td><td colspan="2">0.2</td></tr>
<tr><td rowspan="14">FENet</td><td>decoder_hidden</td><td>512</td><td>256</td></tr>
<tr><td>features_by_layer</td><td colspan="2">[1, 1, 1, 1, 1, 1, 1, 1]</td></tr>
<tr><td>kernel_by_layer</td><td colspan="2">[5, 5, 5, 5, 3, 3, 3]</td></tr>
<tr><td>batch_size</td><td colspan="2">128</td></tr>
<tr><td>calib_epochs</td><td rowspan="3">-</td><td>8</td></tr>
<tr><td>calib_lr</td><td>0.0001</td></tr>
<tr><td>calib_ratio</td><td>0.2</td></tr>
<tr><td>dropout</td><td>0.15</td><td>0.2</td></tr>
<tr><td>epochs</td><td>40</td><td>27</td></tr>
<tr><td>lr</td><td>0.001</td><td>0.00078</td></tr>
<tr><td>max_calib_trials</td><td rowspan="3">-</td><td>30</td></tr>
<tr><td>min_calib_trials</td><td>5</td></tr>
<tr><td>patience</td><td>8</td></tr>
<tr><td>weight_decay</td><td>0.01</td><td>0.02</td></tr>
<tr><td rowspan="9">seqVAE</td><td>beta</td><td colspan="2">1.0</td></tr>
<tr><td>embed_dim</td><td colspan="2">32</td></tr>
<tr><td>encoder_type</td><td colspan="2">transformer</td></tr>
<tr><td>kl_weight</td><td colspan="2">1e-3</td></tr>
<tr><td>latent_dim</td><td colspan="2">24</td></tr>
<tr><td>num_heads</td><td colspan="2">2</td></tr>
<tr><td>num_layers</td><td colspan="2">1</td></tr>
<tr><td colspan="2">prior_hidden</td><td>128</td></tr>
<tr><td>rec_weight</td><td colspan="2">1e-3</td></tr>
</table>

| | batch_size | 128 | |
|---|---|---|---|
| | dropout | 0.3 | 0.35 |
| | epochs | 30 | 40 |
| | lr | 5e-4 | |
| | patience | 8 | 10 |
| | restore_best | True | |
| | save_ckpt | False | |
| | val_ratio | 0.2 | |
| | weight_decay | 1e-4 | |

## Supplementary Table 20. Shared encoder architecture in the LLM–driven AAS.

| Parameters | | Values |
|---|---|---|
| Shared encoder | hidden_dim | 320 |
| | num_layers | 6 |
| | num_heads | 8 |
| | mlp_ratio | 2.5 |
| | dropout | 0.1 |
| | atten_dropout | 0.1 |
| | pre_norm | True |
| | positional | learned |
| | use_task-routing_token | True |
| | ffn_activation | gelu |
| | encoder_layernorm_eps | 1e-5 |
| | weight_init | trunc_normal |

# Supplementary Notes

## Supplementary Note 1. The parameters override those of some selected benchmark decoders across datasets.

**MLP.** For Finger PPC, Reaching C-J-M-T, Reaching B, Jango, Visual Grating, Visual Coding, Music, and Emotion datasets, the following parameter overrides were applied in the classification decoding analysis: *seeds = 44*, *46*, and *54*, and *batch_size=512*.

**LSTM.** For the Jango dataset, the following parameter overrides were applied in the cross-day classification decoding analysis: *batch_size = 128, dropout = 0.3, epochs = 30*, *hidden_size = 64*, *lr = 0.0008*, *num_layers = 3*, *patience = 12*, *pooling = last*, and *weight_decay = 0.0001*. For the Jango dataset, the following parameter overrides were applied for the cross-day regression decoding analysis: *batch_size = 512*; *dropout = 0.3*; *epochs = 30*; *hidden_size = 64*; *lr = 0.0008*; *num_layers = 3*; *weight_decay = 0.0001*. For the FALCON M2 dataset, *best_k_test_days = 3* was set for cross-day regression decoding analysis.

**Transformer.** For the FALCON M2 dataset, parameter overrides were applied in both the single-day and cross-day regression decoding analyses. In the single-day analysis, the overrides were *d_model = 128*, *dropout = 0.2*, *epochs = 60*, *n_heads = 8*, and *num_layers = 3*. In the cross-day analysis, the overrides were *batch_size = 64*, *best_k_test_days = 3*, *calib_epochs = 8*, *calib_lr = 0.0001*, *calib_ratio = 0.2*, *calib_scope = head_encoder*, *d_model = 32*, *dropout = 0.5*, *epochs = 120*, *lr* = $4 \times 10^{-5}$, *max_calib_trials = 30*, *min_calib_trials = 5*, *n_heads = 4*, *num_layers = 1*, *patience = 25*, and *weight_decay = 0.0025*. For the Jango dataset, parameter overrides were applied in both cross-day classification and regression decoding analyses. In the classification analysis, the overrides were *d_model = 128*, *dropout = 0.2*, *epochs = 90*, *n_heads = 8*, *num_layers = 3*, *patience = 20*, and *weight_decay = 0.0001*. In the regression analysis, the overrides were *batch_size = 64*, *calib_epochs = 8*, *calib_lr = 0.0001*, *calib_ratio = 0.2*, *calib_scope = head_encoder*, *d_model = 64*, *dropout = 0.35*, *epochs = 100*, *lr* = $5 \times 10^{-5}$, *max_calib_trials = 30*, *min_calib_trials = 5*, *n_heads = 4*, *num_layers = 2*, *patience = 20*, and *weight_decay = 0.001*. For the LINK CO dataset, the parameter overrides applied in the cross-day regression decoding analysis were *batch_size = 64*, *calib_epochs = 8*, *calib_lr = 0.0001*, *calib_ratio = 0.2*, *calib_scope = head_encoder*, *dropout = 0.3*, *epochs = 90*, *lr* = $6 \times 10^{-5}$, *max_calib_trials = 30*, *min_calib_trials = 5*, *patience = 18*, and *weight_decay = 0.0008*. For the LINK RTT dataset, the parameter overrides applied in the cross-day regression decoding analysis were *batch_size = 64*, *calib_epochs = 15*, *calib_lr = 0.0001*, *calib_ratio = 0.5*, *calib_scope = head_encoder*, *d_model = 64*,

*dropout = 0.35*, *epochs = 90*, *lr* = $5 \times 10^{-5}$, *max_calib_trials = 240*, *min_calib_trials = 5*, *n_heads = 4*, *num_layers = 2*, *patience = 6*, and *weight_decay = 0.001*. For the Indy dataset, the parameter overrides applied in the cross-day regression decoding analysis were *batch_size = 64*, *calib_epochs = 8*, *calib_lr = 0.0001*, *calib_ratio = 0.2*, *calib_scope = head_encoder*, *dropout = 0.25*, *epochs = 90*, *lr* = $7 \times 10^{-5}$, *max_calib_trials = 30*, *min_calib_trials = 5*, *patience = 18*, *weight_decay = 0.0005*. For the LINK CO, LINK RTT, and Indy datasets, the parameter overrides applied in the single-day regression decoding analysis were *dropout = 0.2*, *epochs = 60*, *lr* = $7 \times 10^{-5}$, *patience = 15*, and *weight_decay = 0.0005*.

**MSCFormer.** For the Music dataset, *batch_size=128* in the cross-day classification decoding analysis. For the Jango dataset, the parameter overrides applied in the cross-day classification decoding analysis were *epochs=90*, *batch_size=64*, *lr*=$8 \times 10^{-4}$, *weight_decay*=$1 \times 10^{-4}$, *dropout=0.45*, and *patience=20*. For the Finger PPC, Reaching C-J-M-T, Visual Grating, Visual Coding, and Emotion datasets, *batch_size=64* in the formal cross-day rerun.

**FENet.** For the FALCON M2, Jango, LINK CO, and LINK RTT datasets, the parameter overrides applied in the single-day regression decoding analysis were *decoder_hidden = 512*, *dropout = 0.15*, *epochs = 40*, *lr = 0.001*, and *weight_decay = 0.01*. For the FALCON M2 dataset, the parameter overrides applied in the cross-day regression decoding analysis were *best_k_test_days = 3*, *calib_epochs = 12*, *calib_lr = 0.0001*, *calib_ratio = 0.4*, *calib_scope = head*, *max_calib_trials = 120*, *min_calib_trials = 5*, and *sequence_decoder = global*. For the LINK CO dataset, the parameters override in the cross-day regression decoding analysis were *calib_epochs = 10*, *calib_lr = 0.0001*, *calib_ratio = 0.3*, *calib_scope = head*, *decoder_hidden = 1024*, *dropout = 0.15*, *epochs = 45*, *features_by_layer = [2, 2, 2, 2, 2, 2, 2, 2]*, *lr = 0.0005*, *max_calib_trials = 120*, *min_calib_trials = 5*, *patience = 14*, *sequence_decoder = global*, *weight_decay = 0.01*. For the LINK RTT dataset, the parameter overrides applied in the cross-day regression decoding analysis were *calib_epochs = 12*, *calib_lr = 0.0001*, *calib_ratio = 0.5*, *calib_scope = head*, *decoder_hidden = 1024*, *dropout = 0.2*, *epochs = 45*, *features_by_layer = [2, 2, 2, 2, 2, 2, 2, 2]*, *lr = 0.0004*, *max_calib_trials = 240*, *min_calib_trials = 5*, *patience = 0*, *restore_best = False*, *sequence_decoder = global*, *weight_decay = 0.01*. For the Indy dataset, parameter overrides were applied in both the single-day and cross-day regression decoding analyses. In the single-day analysis, the overrides were *calib_epochs = 12*, *calib_lr = 0.0001*, *calib_ratio = 0.5*, *calib_scope = head_backbone*, *decoder_hidden = 1024*, *dropout = 0.1*, *epochs = 60*, *features_by_layer = [2, 2, 2, 2, 2, 2, 2, 2]*, *lr = 0.0005*, *max_calib_trials = 240*, *min_calib_trials = 5*, *patience = 12*, *restore_best = True*, *sequence_decoder = global*, *weight_decay = 0.005*. In the cross-day analysis, the overrides were *calib_epochs = 12*, *calib_lr = 0.0001*, *calib_ratio = 0.5*, *calib_scope = head_backbone*, *decoder_hidden = 1024*, *dropout = 0.1*, *epochs = 60*, *features_by_layer = [2, 2, 2, 2, 2, 2, 2, 2]*, *lr = 0.0005*, *max_calib_trials = 240*, *min_calib_trials = 5*, *patience = 12*,

*restore_best = True*, *sequence_decoder = global*, *weight_decay = 0.005*.

**DFINE.** For the FALCON M2 dataset, parameter overrides were applied in both the single-day and cross-day regression decoding analyses. In the single-day analysis, the overrides were *hidden_layer_list = [32, 32, 32]*, *hidden_layer_list_mapper = [32, 32]*, *lr = 0.01*, *scale_behv_recons = 10*, and *steps_ahead = [1, 2, 3]*. In the cross-day analysis, the overrides were *best_k_test_days = 3*, *calib_epochs = 80*, *calib_lr = 0.0003*, *calib_ratio = 0.5*, *calib_scope = mapper*, *dim_a = 4*, *dim_x = 4*, *hidden_layer_list = [32, 32, 32]*, *hidden_layer_list_mapper = [32, 32]*, *max_calib_trials = 240*, *min_calib_trials = 5*, and *scale_behv_recons = 30*. For the Jango and Indy datasets, the parameter overrides applied in the single-day regression decoding analysis were *hidden_layer_list = [20, 20, 20, 20]*, *hidden_layer_list_mapper = [20, 20, 20]*, *lr = 0.02*, *scale_behv_recons = 20*, and *steps_ahead = [1, 2, 3, 4]*. For the Jango dataset, the parameters override in the cross-day regression decoding analysis were *calib_epochs = 20*, *calib_lr = 0.0001*, *calib_ratio = 0.3*, *calib_scope = mapper*, *dim_a = 4*, *dim_x = 4*, *epochs = 90*, *grad_clip = 0.75*, *hidden_layer_list = [48, 48, 48]*, *hidden_layer_list_mapper = [48, 48]*, *lr = 0.007*, *max_calib_trials = 120*, *min_calib_trials = 5*, *scale_behv_recons = 12*, and *steps_ahead = [1, 2, 3]*. For the LINK CO dataset, parameter overrides were applied in both the single-day and cross-day regression decoding analyses. In the single-day analysis, the overrides were *calib_epochs = 20*, *calib_lr = 0.0001*, *calib_ratio = 0.3*, *calib_scope = mapper*, *dim_a = 4*, *dim_x = 4*, *epochs = 90*, *grad_clip = 0.75*, *hidden_layer_list = [48, 48]*, *hidden_layer_list_mapper = [48, 48]*, *lr = 0.007*, *max_calib_trials = 120*, *min_calib_trials = 5*, *scale_behv_recons = 12*, and *steps_ahead = [1, 2, 3]*. In the cross-day analysis, the overrides were *calib_epochs = 40*, *calib_lr = 0.0003*, *calib_ratio = 0.5*, *calib_scope = mapper*, *max_calib_trials = 240*, and *min_calib_trials = 5*. For the LINK RTT dataset, the parameter overrides applied in the cross-day regression decoding analysis were *calib_epochs = 80*, *calib_lr = 0.001*, *calib_ratio = 0.5*, *calib_scope = mapper*, *dim_a = 4*, *dim_x = 4*, *hidden_layer_list = [32, 32]*, *hidden_layer_list_mapper = [32, 32]*, *max_calib_trials = 240*, *min_calib_trials = 5*, and *scale_behv_recons = 30*.

**LFADS.** For the FALCON M2 dataset, the parameter overrides applied in the single-day regression decoding analysis were *epochs = 80*, *lr = 0.003*, *patience = 15*. For the 5_Jango_force dataset, parameter overrides were applied in the following analyses. In the single-day regression analysis, the overrides were *batch_size = 128*, *decoder_type = mlp*, *dropout = 0.1*, *epochs = 80*, *fac_dim = 16*, *lr = 0.001*, *mlp_alpha = 0.0001*, *mlp_hidden_sizes = [128, 64]*, *mlp_max_iter = 400*, *patience = 15*. In the cross-day regression analysis, the overrides were *batch_size = 256*, *decoder_type = mlp*, *dropout = 0.1*, *epochs = 80*, *fac_dim = 16*, *lr = 0.001*, *mlp_hidden_sizes = [128, 64]*, and *patience = 15*. In the cross-day classification analysis, the overrides were *batch_size = 64*, *dropout = 0.02*, *epochs = 80*, *fac_dim = 40*, *lr = 0.003*, and *patience = 15*. For the LINK CO dataset, the parameter overrides applied in the single-day regression decoding analysis were *batch_size = 256*, *decoder_type = ridge*, *dropout = 0.02*, *epochs = 180*, *fac_dim = 40*, *lr = 0.0015*, *patience = 35*, *reduce = sequence*, and *val_ratio = 0.1*. For the LINK RTT dataset, the parameter overrides applied in the single-day regression decoding analysis were *batch_size = 256*, *decoder_type = ridge*, *dropout = 0*, *epochs = 140*, *fac_dim = 40*, *lr = 0.002*, *patience = 30*, *reduce = sequence*, and *val_ratio = 0.1*. For the FALCON M2, LINK CO, and LINK RTT datasets, the parameter overrides applied in the cross-day regression decoding analysis were *batch_size = 256*, *dropout = 0.1*, *epochs = 80*, *fac_dim = 16*, *lr = 0.001*, and *patience = 15*. For the Indy dataset, parameter overrides were applied in both the single-day and cross-day regression decoding analyses. In the single-day analysis, the overrides were *batch_size = 128*, *decoder_type = mlp*, *dropout = 0.1*, *epochs = 80*, *fac_dim = 16*, *lr = 0.001*, *mlp_alpha = 0.0001*, *mlp_hidden_sizes = [128, 64]*, *mlp_max_iter = 400*, *patience = 15*. In the cross-day analysis, the overrides were *batch_size = 256*, *decoder_type = ridge*, *dropout = 0.1*, *epochs = 70*, *fac_dim = 32*, *lr = 0.001*, and *patience = 12*. For the Visual Grating and Music datasets, the parameter overrides applied in the cross-day classification decoding analysis were *batch_size = 32*, *dropout = 0.1*, *epochs = 60*, *fac_dim = 24*, *lr = 0.0005*, *patience = 15*, and *weight_decay = 0.0001*.

**Stabilization.** For the FALCON M2 dataset, parameter overrides were applied in both the single-day and cross-day regression decoding analyses. In the single-day analysis, the overrides were *align_n = 20*, *align_th = 0.001*, *baseline_ntrials = 1000000000*, *best_k_test_days = 3*, *calib_ratio = 0*, *decoder_type = kalman*, *fa_min_priv_var = 0.1*, *fine_tune = False*, *fine_tune_epochs = 100*, *fine_tune_include_train = False*, *fit_state_noise = True*, *hidden_sizes = [64, 32]*, *max_calib_trials = 0*, *min_calib_trials = 0*, *n_latents = 10*, and *single_day_mode = freeze_base*. In the cross-day analysis, the overrides were *align_n = 20*, *align_th = 0.05*, *baseline_ntrials = 80*, *best_k_test_days = 3*, *calib_ratio = 0.2*, *decoder_type = kalman*, *fa_min_priv_var = 0.1*, *fine_tune = False*, *fit_state_noise = True*, *hidden_sizes = [64, 32]*, *max_calib_trials = 30*, *min_calib_trials = 5*, and *n_latents = 10*. For the LINK CO dataset, parameter overrides were applied in both the single-day and cross-day regression decoding analyses. In the single-day analysis, the overrides were *align_n = 20*, *align_th = 0.001*, *baseline_ntrials = 1000000000*, *calib_ratio = 0*, *decoder_type = kalman*, *fa_min_priv_var = 0.5*, *fine_tune = False*, *fine_tune_epochs = 100*, *fine_tune_include_train = False*, *fit_state_noise = True*, *hidden_sizes = [64, 32]*, *max_calib_trials = 0*, *min_calib_trials = 0*, *n_latents = 8*, and *single_day_mode = freeze_base*. In the cross-day analysis, the overrides were *align_n = 20*, *align_th = 0.001*, *baseline_ntrials = 120*, *calib_ratio = 0.5*, *decoder_type = kalman*, *fa_min_priv_var = 0.5*, *fine_tune = True*, *fine_tune_include_train = False*, *fit_state_noise = True*, *hidden_sizes = [64, 32]*, *max_calib_trials = 360*, *min_calib_trials = 5*, and *n_latents = 10*. For the LINK RTT dataset, parameter overrides were applied in both the single-day and cross-day regression decoding analyses. In the single-day analysis, the overrides were *align_n = 60*, *align_th = 0.01*, *baseline_ntrials = 1000000000*, *calib_ratio = 0*, *decoder_type = kalman*,

*fa_min_priv_var = 0.1*, *fine_tune = False*, *fine_tune_epochs = 100*, *fine_tune_include_train = False*, *fit_state_noise = True*, *max_calib_trials = 0*, *min_calib_trials = 0*, *n_latents = 12*, and *single_day_mode = freeze_base*. In the cross-day analysis, the overrides were *calib_ratio = 0.5*, *fine_tune_include_train = False*, *max_calib_trials = 240*, and *min_calib_trials = 5*. For the Indy dataset, parameter overrides were applied in both the single-day and cross-day regression decoding analyses. In the single-day analysis, the overrides were *align_n = 60*, *align_th = 0.01*, *calib_ratio = 0.5*, *fine_tune = True*, *fine_tune_epochs = 25*, *fine_tune_include_train = False*, *max_calib_trials = 240*, and *min_calib_trials = 5*. In the cross-day analysis, the overrides were *calib_ratio = 0.5*, *fine_tune_epochs = 25*, *fine_tune_include_train = False*, *max_calib_trials = 240*, and *min_calib_trials = 5*.

**Cycle-GAN.** With respect to the FALCON M2 dataset, the parameters applied in the single-day regression decoding analysis were *calib_scope = decoder*, *decoder_epochs = 40*, *single_day_mode = decoder_only*, and *source_ratio = 0.98*. For the Jango dataset, parameter overrides were applied in the following decoding analyses. In the single-day regression analysis, the overrides were *calib_scope = aligner*, *cycle_weight = 5*, *decoder_epochs = 80*, *decoder_type = mlp*, *epochs = 500*, *identity_weight = 5*, *max_calib_trials = 28*, *single_day_mode = decoder_only*, and *source_ratio = 0.95*. In the cross-day regression analysis, the overrides were *cycle_weight = 7.5*, *decoder_epochs = 50*, *epochs = 500*, and *identity_weight = 7.5*. In the cross-day classification analysis, the overrides were *calib_ratio = 0.15*, *cycle_weight = 7.5*, *decoder_epochs = 50*, *epochs = 500*, *identity_weight = 7.5*, *max_calib_trials = 28*, and *min_calib_trials = 15*. For the LINK-CO dataset, the parameter overrides applied in the single-day regression decoding analysis were *calib_epochs = 12*, *calib_lr = 0.0001*, *calib_scope = decoder*, *decoder_epochs = 40*, *decoder_hidden = [128]*, *decoder_num_layers = 2*, *decoder_type = lstm*, *dropout_decoder = 0.2*, *single_day_mode = decoder_only*, and *source_ratio = 0.98*. For the LINK RTT dataset, the parameter overrides applied in the single-day regression decoding analysis were *calib_epochs = 12*, *calib_lr = 0.0001*, *calib_scope = all*, *decoder_epochs = 40*, *decoder_hidden = [128]*, *decoder_num_layers = 2*, *decoder_type = lstm*, and *dropout_decoder = 0.2*. For the Indy dataset, the parameter overrides applied in the single-day regression decoding analysis were *cycle_weight = 5*, *epochs = 120*, and *identity_weight = 5*.

**NoMAD.** For the Jango dataset, the parameter overrides when applied in both cross-day classification and regression decoding analysis. In the classification analysis, the overrides were *epochs=80*, *batch_size=64*, *align_epochs=5*, and *patience=15*. In the regression analysis, the override was *align_epochs=5*. For the Reaching B and Music datasets, *batch_size=128* in the cross-day classification decoding analysis. For the Finger PPC, Reaching C-J-M-T, Visual Rating, Visual Coding, and Emotion datasets, *batch_size=64* in the formal cross-day rerun.

**seqVAE.** Regression runs on FALCON M2, LINK CO, LINK RTT, and Indy use seed=44 and write both single-day and per-test-day cross-day entries.

**MFSNN.** For the Reaching B and Music datasets, the parameter overrides applied in the cross-day classification decoding analysis were *batch_size=128* and *eval_interval=5*. For the Jango dataset, the parameter overrides applied in the cross-day classification decoding analysis were *batch_size=64*, *normalize=zscore_per_trial*, *epochs=100*, *steps=20*, and *patience=20*. For the Finger PPC, which reached the C-J-M-T, Visual Grating, Visual Coding, and Emotion datasets, the parameter overrides applied in the formal cross-day rerun were *batch_size=64* and *eval_interval=5*.